\documentclass[aps,physrev,reprint,onecolumn,amssymb,amsmath,superscriptaddress,showkeys]{revtex4-2}

\usepackage{tikz}
\usetikzlibrary{arrows, shapes, patterns, decorations.pathmorphing, arrows.meta}

\usepackage{graphicx}
\usepackage[normalem]{ulem}
\usepackage{cases}
\usepackage{bbm}
\usepackage{xcolor}

\usepackage{hyperref}
\hypersetup{colorlinks=true,
linkcolor=purple,
urlcolor=teal,
citecolor=violet}

\newcommand{\e}{\mathrm{e}}
\newcommand{\dd}{\mathrm{d}}

\begin{document}

\title{Effect of slow bonds on current fluctuations in the symmetric simple exclusion process}

\author{Soumyabrata Saha}
\email{soumyabrata.saha@tifr.res.in}
\affiliation{Department of Theoretical Physics, Tata Institute of Fundamental Research, Homi Bhabha Road, Mumbai 400005, India}

\author{Sandeep Jangid}
\email{sandeep.jangid@tifr.res.in}
\affiliation{Department of Theoretical Physics, Tata Institute of Fundamental Research, Homi Bhabha Road, Mumbai 400005, India}

\author{Kapil Sharma}
\email{kapil.sharma@tifr.res.in}
\affiliation{Department of Theoretical Physics, Tata Institute of Fundamental Research, Homi Bhabha Road, Mumbai 400005, India}

\author{Tridib Sadhu}
\email{tridib@theory.tifr.res.in}
\affiliation{Department of Theoretical Physics, Tata Institute of Fundamental Research, Homi Bhabha Road, Mumbai 400005, India}

\date{\today}

\begin{abstract}
The symmetric simple exclusion process (SSEP) is a paradigmatic model of classical non-equilibrium dynamics. Exact results for large deviations of particle current in the SSEP have been obtained in various settings using integrability-based methods. In this Article, we discuss how these results are modified in the presence of localized slow bonds. We consider three conventional geometries: (a) a finite one-dimensional lattice weakly coupled to unequal reservoirs at its boundaries, (b) a semi-infinite one-dimensional lattice weakly coupled to a boundary reservoir, and (c) an infinite one-dimensional lattice with localized slow bonds near the origin. For each case, we present exact expressions for the large deviation function of current and validate them through rare-event simulations based on the cloning algorithm. In connection with our results, we present an elementary derivation of the exact large deviation function for the current in the semi-infinite SSEP, complementing recent results obtained through more elaborate techniques.
\end{abstract}

\keywords{symmetric simple exclusion process, fluctuating hydrodynamics, macroscopic fluctuation theory, large deviations of current, slow bond}

\maketitle

\tableofcontents

\section{Introduction}
Out-of-equilibrium systems are ubiquitous in nature, exhibiting a rich array of emergent phenomena that distinguish them from their equilibrium counterparts. Understanding these systems requires developing a statistical framework for non-equilibrium fluctuations, analogous to equilibrium statistical mechanics. 
Much of our current understanding of fluctuations in extended non-equilibrium systems arises from the study of analytically tractable models of low-dimensional interacting particle systems with irreversible stochastic dynamics \cite{1991_Spohn_Large,1999_Kipnis_Scaling,1999_Liggett_Stochastic}. Among these, the Symmetric Simple Exclusion Process (SSEP) stands out as one of the simplest and extensively studied models \cite{2025_Derrida_Les_Houches,2007_Derrida_Non,2015_Mallick_The}. Over the past decades, the SSEP and its variants have become paradigmatic in non-equilibrium statistical mechanics, demonstrating broad applicability across diverse domains, from biological transport \cite{2011_Chou_Non} to chaotic quantum dynamics \cite{2023_McCulloch_Full,turkeshi2024,2026_mcculloch}.

Among the most prominent exact results about fluctuations in the SSEP are the full counting statistics of particle flux in different geometries \cite{2025_Derrida_Les_Houches,2007_Derrida_Non,2015_Mallick_The}, each highlighting distinct aspects of non-equilibrium fluctuations.

The first result pertains to the SSEP on a finite one-dimensional lattice coupled at the boundaries to particle reservoirs with unequal densities. Over time, the system reaches a non-equilibrium steady state (NESS), where the net number of particles transferred between the reservoirs follows a large deviation statistics. An exact expression of the corresponding large deviation function has been independently derived using a variational principle \cite{2004_Bodineau_Current,2007_Bodineau_Cumulants} and by diagonalizing the tilted Markov generator \cite{2004_Derrida_Current}.

The second result concerns non-stationary fluctuations in the SSEP on an infinite one-dimensional lattice with a domain-wall initial condition. As the system evolves towards an asymptotic equilibrium state, the large deviation statistics of the net particle flux between the two halves of the system has been obtained using an ingenious application of the Bethe ansatz technique \cite{2009_Derrida_Current}.

The third result involves the SSEP on a semi-infinite one-dimensional lattice coupled to a single particle reservoir at the origin. Here, the system evolves towards an asymptotic equilibrium state with the reservoir. At long times, the net particle flux from the reservoir into the system follows a large deviation statistics, recently derived \cite{2026_Sharma_Large} (see also~\cite{2024_Grabsch}) by invoking a mapping to the infinite-lattice SSEP. Notably, for both the infinite and semi-infinite cases, non-stationary current fluctuations exhibit slow relaxation, where fluctuations remain sensitive to the initial state even at long times \cite{2009_Derrida_Current2,2023_Saha_Current}.

While the SSEP on a regular lattice provides a rich set of exact results, real-world systems often feature local disorder due to impurities, structural defects, or external perturbations. In equilibrium systems, local disorder typically does not affect large-scale behaviour of static observables, where the response to local perturbations is local due to short range correlations. However, in out-of-equilibrium systems, their effect can be non-local even for static quantities \cite{2011_Sadhu_Long,2014_Sadhu_Long,Sadhu2014NonLocal,Maes2009Nonequilibrium}. For dynamic quantities, such as current, localized microscopic disorder can significantly alter their statistics even at equilibrium, raising questions about robustness of earlier seminal results about large fluctuations in realistic scenarios. This makes it necessary to closely examine the effect of local disorder in tractable model systems.

In stochastic lattice gases like the SSEP, local disorder can be introduced in several ways. One common example is a localized defect \cite{1992_Janowsky_Finite,1993_Schütz_Generalized,1994_Janowsky_Exact}, where hopping across a specific bond is significantly slowed down. Alternatively, a localized drive may impose an asymmetric bias on hopping dynamics along a particular bond \cite{2011_Sadhu_Long,2014_Sadhu_Long}. Such defects or drives can also be positioned at lattice sites rather than bonds \cite{2006_Juhász_Partially,2015_Bahadoran_Properties,2016_Franco_Scaling,2025_Sakai_Unexpected}, further enriching the system’s complexity. Another scenario involves incorporating second class particles, like a driven tracer \cite{2017_Cividini_Driven,2020_Lobaskin_Driven,2023_Grabsch_Driven}, whose microscopic dynamics differ from those of the surrounding bath particles.

These various forms of locally disordered systems have primarily been studied using exact microscopic techniques \cite{2022_Lobaskin_Matrix,2023_Lobaskin_Integrability}. However, these techniques are often cumbersome and apply only to specially curated scenarios of simple model dynamics. Even in the absence of local disorder, exact results on non-equilibrium fluctuations remain limited \cite{2025_Derrida_Les_Houches,2007_Derrida_Non,2015_Mallick_The}. This highlights the need for a simpler yet broadly applicable framework, one that not only recovers known exact results, particularly in the large scale limit, but also provides insights into previously unexplored dynamics.

A powerful framework along this line is the macroscopic fluctuation theory (MFT) \cite{2002_Bertini_Macroscopic,2007_Derrida_Non,2015_Bertini_Macroscopic}, which is built on a fluctuating hydrodynamics approach. The MFT has successfully reproduced large deviation results for the current in the SSEP on both finite \cite{2004_Bodineau_Current,2024_Saha_Large} and infinite lattices \cite{2024_Mallick_Exact,2022_Mallick_Exact}, revealing remarkable connections with classical integrability. More recently, it has enabled the derivation of corresponding large deviations result for the semi-infinite geometry \cite{2024_Grabsch, 2026_Sharma_Large}, which was difficult through microscopic techniques.
Beyond the SSEP, the hydrodynamic framework of the MFT has been widely applied to various other well-known \cite{2026_saha_density,2015_Bertini_Macroscopic} stochastic lattice models, including the zero-range processes \cite{2005_Evans_Nonequilibrium,2005_Harris_Current}, the weakly asymmetric simple exclusion process (WASEP) \cite{2004_Enaud_Large,2012_Gorissen_Current}, the Kipnis-Marchioro-Presutti (KMP) model \cite{1982_Kipnis_Heat,2005_Bertini_Large}, the symmetric simple partial exclusion process (SSPEP)\cite{1994_Schütz_Non,2023_Franceschini_Hydrodynamical,2026_Saha}, the symmetric simple inclusion process (SSIP) \cite{2007_Giardinà_Duality,2014_Vafayi_Weakly}, the symmetric simple double exclusion process (SSDEP) \cite{2001_Hager_Minimal,2013_Krapivsky_Dynamics,2017_Baek_Dynamical,2026_Saha}, and the Katz-Lebowitz-Spohn (KLS) model \cite{1983_Katz_Phase,1984_Katz_Nonequilibrium}, many of which are not tractable by microscopic integrability techniques. It has also proven useful in studies of continuum models such as the random average processes \cite{2016_Kundu_Exact,2016_Cividini_Correlation}, the Brownian hard rods \cite{1991_Spohn_Large,2026_Grabsch_Macroscopic,2026_saha_tracer}, and even systems with active \cite{2023_Agranov_Macroscopic,2025_Mukherjee_Hydrodynamics} and Hamiltonian dynamics \cite{2023_Doyon_Ballistic,2026_Kethepalli_Ballistic,2026_saha_tracer}.

The framework of the MFT has so far largely been applied in the absence of local disorder. In fact, incorporating microscopic local disorder into this macroscopic theory presents a non-trivial challenge, as it is not immediate how they would stay relevant after coarse-graining. This question has drawn increasing attention in recent years. In \cite{2017_Baldasso_Exclusion,2020_Erignoux_HydrodynamicsI,2020_Erignoux_HydrodynamicsII}, the noiseless hydrodynamic description was rigorously derived for the SSEP on a finite lattice coupled to the boundary reservoirs via defect bonds with slow jump rates. Later the typical Gaussian fluctuations complementing the average hydrodynamic evolution was obtained in \cite{2017_Franco_Equilibrium,2008_Landim_Stationary,2019_Franco_Non,2020_Goncalves_Non} and the dynamical large deviations was rigorously studied in \cite{2022_Franco_Dynamical,2023_Franco_Large}. Some of these results about hydrodynamics have been extended to SSEP with multiple slow bonds \cite{2013_Franco_Hydrodynamical}, and to the weakly asymmetric exclusion process \cite{2021_Capitão_Hydrodynamics} and the symmetric simple inclusion process \cite{2022_Franceschini_Symmetric} on a finite lattice with slow coupling to reservoirs. A related scenario, where the slow bond is inside the bulk of the lattice for a SSEP coupled between two boundary reservoirs, the typical Gaussian fluctuations of the integrated current and tracer position, and the existence of a large deviations principle were reported in \cite{2013_Franco_Phase} and \cite{2017_Franco_Large}, respectively. 

Relatively, there are much limited \emph{explicit} results about effect of disorder on macroscopic fluctuations in terms of large deviations. For the finite SSEP slowly coupled between reservoirs, the exact large deviation function of the steady state density fluctuations were obtained using microscopic techniques \cite{2021_Derrida_Large} and later recovered \cite{2024_Saha_Large,Bouley2025SteadyState} by using fluctuating hydrodynamics, generalizing earlier seminal works on boundary driven SSEP \cite{2001_Derrida_Free,2002_Bertini_Macroscopic,2004_Bodineau_Current,2007_Bodineau_Cumulants}. These results particularly demonstrate the robustness of macroscopic fluctuations in NESS against boundary perturbations, an aspect crucial for extending the concept of thermodynamic state outside equilibrium.

In this article, we focus on obtaining \emph{explicit} results about large deviations of current in presence of localized defect bonds, quantifying their effect on transport properties in the SSEP by suitably extending the hydrodynamic framework of MFT. For this purpose, we consider all three standard geometries that have appeared in earlier seminal works: the SSEP on an infinite lattice, on a semi-infinite lattice, and on a finite lattice. For each geometry, we present exact expressions for the corresponding large deviation function of the current. Our analysis is based on a new coarse-graining scheme that constructs fluctuating hydrodynamics description in a systematic manner starting directly from the stochastic microscopic dynamics. This construction shows how the macroscopic approach of MFT can be extended to account for the influence of microscopic localized defect bonds.

We have organized this article in the following order. In Sec.~\ref{sec: SOR}, we present a summary of the main results about large deviations of current, including their numerical confirmation using rare-event simulations. In Sec.~\ref{sec:inf SSEP} we discuss in details how the current statistics is obtained using the hydrodynamics approach of MFT for the SSEP on an infinite lattice with a single slow bond. In the following two sections \ref{sec:si full} and \ref{sec:fin full} we briefly discuss the corresponding analysis for the semi-infinite and finite lattice, respectively. In Sec.~\ref{sec:fhd full} we present how the fluctuating hydrodynamics equation is modified in presence of slow bonds for the three geometries of SSEP. We conclude in Sec.~\ref{sec:conclusion} with comments on potential impacts of our results.

\section{The model and a summary of results \label{sec: SOR}}

The SSEP \cite{2007_Derrida_Non,2011_Chou_Non} refers to a stochastic model of hard-core particles performing continuous-time unbiased random walk on a lattice with on-site hard-core interactions. The interaction excludes any two particles occupying the same site of the lattice at any given moment. The configuration of the system at a time $\tau$ is specified by the occupation variables, $n_i(\tau)$, taking values $0$ or $1$ depending on whether the site $i$ is occupied or empty. The system evolves in time following a Markovian dynamics, where a particle hops across a bond without a preferred direction, provided the target site is vacant. We are interested in the examples where the hopping rate is uniform and set to unity across majority of the bonds of the lattice. There are localized defect bonds on the lattice across which the hopping rate is $\gamma$. 

\subsection{SSEP on an infinite lattice with localized defect bonds}

\begin{figure}[t]
\centering
\resizebox{\linewidth}{!}
{
\begin{tikzpicture}
\node [circle, draw=black, thick, fill=blue, inner sep=0pt, minimum size=15pt] (1) at (1,0) {\footnotesize\color{white}-7};
\node [circle, draw=black, thick, fill=blue, inner sep=0pt, minimum size=15pt] (7) at (7,0) {\footnotesize\color{white}-1};
\node [circle, draw=black, thick, fill=white, inner sep=0pt, minimum size=15pt] (10) at (10,0) {\footnotesize2};
\node [circle, draw=black, thick, fill=white, inner sep=0pt, minimum size=15pt] (16) at (16,0) {\footnotesize8};

\draw[ultra thick, dashed] (1) -- (0,0);
\draw[thick] (1) -- (7);
\draw[thick, decorate, decoration={zigzag, segment length=2.5pt, amplitude=1.25pt}] (7) -- (10);
\draw[thick] (10) -- (16);
\draw[ultra thick, dashed] (16) -- (17,0);

\node [circle, draw=black, thick, fill=blue, inner sep=0pt, minimum size=15pt] (2) at (2,0) {\footnotesize\color{white}-6};
\node [circle, draw=black, thick, fill=white, inner sep=0pt, minimum size=15pt] (3) at (3,0) {\footnotesize-5};
\node [circle, draw=black, thick, fill=white, inner sep=0pt, minimum size=15pt] (4) at (4,0) {\footnotesize-4};
\node [circle, draw=black, thick, fill=blue, inner sep=0pt, minimum size=15pt] (5) at (5,0) {\footnotesize\color{white}-3};
\node [circle, draw=black, thick, fill=white, inner sep=0pt, minimum size=15pt] (6) at (6,0) {\footnotesize-2};

\node [circle, draw=black, thick, fill=white, inner sep=0pt, minimum size=15pt] (8) at (8,0) {\footnotesize0};
\node [circle, draw=black, thick, fill=blue, inner sep=0pt, minimum size=15pt] (9) at (9,0) {\color{white}\footnotesize1};
\node [circle, draw=black, thick, fill=blue, inner sep=0pt, minimum size=15pt] (11) at (11,0) {\color{white}\footnotesize3};
\node [circle, draw=black, thick, fill=white, inner sep=0pt, minimum size=15pt] (12) at (12,0) {\footnotesize4};
\node [circle, draw=black, thick, fill=blue, inner sep=0pt, minimum size=15pt] (13) at (13,0) {\color{white}\footnotesize5};
\node [circle, draw=black, thick, fill=blue, inner sep=0pt, minimum size=15pt] (14) at (14,0) {\color{white}\footnotesize6};
\node [circle, draw=black, thick, fill=blue, inner sep=0pt, minimum size=15pt] (15) at (15,0) {\color{white}\footnotesize7};

\draw [-{Triangle[width=2.5mm, length=2.5mm]}, very thick] (2.north) to [out=60,in=120] node [midway, above]{\Large$1$} (3.north);
\draw [-{Triangle[width=2.5mm, length=2.5mm]}, very thick] (5.south) to [out=-120,in=-60] node [midway, below]{\Large$1$} (4.south);
\draw [-{Triangle[width=2.5mm, length=2.5mm]}, very thick] (5.north) to [out=60,in=120] node [midway, above]{\Large$1$} (6.north);
\draw [-{Triangle[width=2.5mm, length=2.5mm]}, very thick] (7.south) to [out=-120,in=-60] node [midway, below]{\Large$1$} (6.south);
\draw [-{Stealth[width=3.0mm, length=3.0mm]}, ultra thick] (11.south) to [out=-120,in=-60] node [midway, below]{\Large$1$} (10.south);
\draw [-{Triangle[width=2.5mm, length=2.5mm]}, very thick] (11.north) to [out=60,in=120] node [midway, above]{\Large$1$} (12.north);
\draw [-{Triangle[width=2.5mm, length=2.5mm]}, very thick] (13.south) to [out=-120,in=-60] node [midway, below]{\Large$1$} (12.south);
\draw [-{Triangle[width=2.5mm, length=2.5mm]}, very thick] (15.north) to [out=60,in=120] node [midway, above]{\Large$1$} (16.north);

\draw [-{Stealth[width=4.0mm, length=4.0mm]}, very thick] (7.north) to [out=60,in=135] node [midway, above]{\Large$\gamma$} (8.north);
\draw [-{Stealth[width=4.0mm, length=4.0mm]}, very thick] (9.south) to [out=-120,in=-45] node [midway, below]{\Large$\gamma$} (8.south);
\draw [-{Stealth[width=4.0mm, length=4.0mm]}, ultra thick] (9.north) to [out=60,in=135] node [midway, above]{\Large$\gamma$} (10.north);
\end{tikzpicture}
}
\caption{\textbf{Infinite-line SSEP:} The SSEP on an infinite integer lattice with $\ell$ localized defect bonds (zigzag line) placed around the bond between sites $0$ and $1$. Symmetric jump rates across the bonds are indicated, which is $\gamma$ for the defect bonds and $1$ for the rest. The schematic shows an example with three defect bonds placed around the origin. The empty circles indicate unoccupied sites while the filled circles indicate sites occupied by a particle.}
\label{fig:inf_with_l_slow_bonds}
\end{figure}
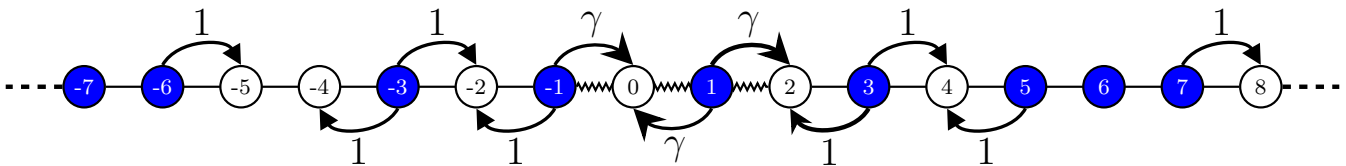

Sites on the one-dimensional integer lattice is indexed by $i\equiv\{\cdots,-2,-1,0,1,2,\cdots\}$, where $\ell$-number of defect bonds are localized around the origin. A schematic of the geometry is shown in Fig.~\ref{fig:inf_with_l_slow_bonds}.
The system is initially prepared in a domain wall state where, negative sites ($i < 0$) are filled by particles following a Bernoulli measure with a uniform average density $\rho_a$, while the positive sites ($i\ge 0$) are accordingly filled with a uniform average density $\rho_b$. 

As the system evolves, the two halves exchange particles in an attempt to reach an eventual equilibrium state with each other at an intermediate density. There is a net average flux from the dense half to the lighter half. Our interest is in the net flux of particles $Q_T$ across the bond between $0$-th and the $1$-th sites measured over a large time window $T$. The statistics of $Q_T$ is governed by both the fluctuations in the initial states as well as the stochastic evolution of the particles.

The problem in absence of defect bonds ($\gamma=1$) has been extensively studied in \cite{2009_Derrida_Current,2009_Derrida_Current2,2022_Mallick_Exact,2024_Mallick_Exact}. It was rigorously shown \cite{2009_Derrida_Current} that all cumulants of $Q_T$ scales as $\sqrt{T}$ for large time $T$. Equivalently, the generating function
\begin{equation}\label{eq:generating function infinite}
\big<\e^{\lambda Q_T}\big>\asymp e^{\sqrt{T}\,\mu_{\text{inf}}(\lambda,\rho_a,\rho_b)}
\end{equation}
at large $T$, where the scaled cumulant generating function (scgf), $\mu_{\text{inf}}$, depends on the parameters $(\lambda, \rho_a, \rho_b)$ through a single function 
\begin{equation}\label{omega_param_defn}
\omega(\lambda,\rho_a,\rho_b)=(\e^\lambda-1)\rho_a(1-\rho_b)+(\e^{-\lambda}-1)\rho_b(1-\rho_a)
\end{equation}
with an explicit expression \cite{2009_Derrida_Current}
\begin{equation}\label{eq:mu fast inf}
\mu_{\text{inf}}(\lambda,\rho_a,\rho_b)=R_{\text{inf}}\big(\omega(\lambda,\rho_a,\rho_b)\big)\quad \textrm{and }R_{\text{inf}}(\omega)=\int_{-\infty}^\infty\frac{\dd k}{\pi}\log{\big(1+\omega\e^{-k^2}\big)}.
\end{equation}
The dependence on $\omega$ is a remarkable consequence of the rotational symmetry \cite{2009_Derrida_Current2,2008_Appert_Universal} of the generator of the dynamics of SSEP. The symbol $\asymp$ in \eqref{eq:generating function infinite} means that the ratio of logarithm of two sides converge to $1$ for $T\to \infty$.

The solution has alternative representations, indicating potential connections to other problems. For example, $R_{\text{inf}}(\omega)=-\frac{1}{\sqrt{\pi}}\mathrm{Li}_{3/2}(-\omega)$ in terms of the poly-logarithm function $\mathrm{Li}_s(z)$ has similarity with the expression of the number of particles in Fermi-Dirac statistics and with the scgf for the number of surviving particles in an assembly of annihilating random walkers \cite{2001_Schutz_Exactly}. Another representation in terms of Fredholm determinant
$R(\omega)=\frac{1}{\pi}\log{\det{[\mathbbm{1}+\omega K(x,y)]}}$
with the kernel $K(x,y)=\delta(x-y)\exp{(\partial_x^2)}$, finds analogue in the current statistics for asymmetric exclusion process~\cite{2000_Johansson_Shape,2008_Tracy_A,2009_Tracy_Total}.

In presence of fast defect bonds ($\gamma>1$) it is expected that the result (\ref{eq:generating function infinite}-\ref{eq:mu fast inf}) remains unchanged as the bottleneck for transport comes from the bulk of the lattice. On the other hand, for slow defect bonds ($\gamma<1$), the defects would serve as a bottleneck, and could significantly modify the statistics. However, it turns out that this is not entirely correct. We show that large time statistics of $Q_T$ in \eqref{eq:generating function infinite} remains robust for the broad range of jump rates $\gamma$ larger than $\sim T^{-1/2}$ when the number $\ell$ of defect bonds is finite. Only when the rate is extremely small $\gamma\sim T^{-1/2}$ or smaller, the slow bonds act as a bottleneck and affect the long time statistics of $Q_T$.

This result can be explicitly demonstrated by studying the marginal case 
$\gamma=\Gamma/\sqrt{T}$ with $\Gamma$ a positive valued parameter. We show that the long-time statistics of $Q_T$ still obeys the asymptotics \eqref{eq:generating function infinite}, but with a modified scgf $\mu_{\text{inf}}^{\text{slow}}$ that explicitly depends on the parameter $\Gamma$ and the number $\ell$ of defect bonds. The dependence on parameters $(\lambda,\rho_a,\rho_b)$ remains through the function $\omega$ in \eqref{omega_param_defn} with 
\begin{equation}\label{eq:mu_slow_inf}
\mu_{\text{inf}}^{\text{slow}}(\lambda,\rho_a,\rho_b,\Gamma,\ell)=R_{\text{inf}}^{\text{slow}}\big(\omega(\lambda,\rho_a,\rho_b),\Gamma,\ell \big),
\end{equation}
where $R_{\text{inf}}^{\text{slow}}$ has a variational formula:
\begin{equation}\label{eq:variational infinite ssep}
R_{\text{inf}}^{\text{slow}}(\omega,\Gamma,\ell)=\min_{z_a, z_b}\Bigg\{R_{\text{si}}\big(\sinh^2{z_a}\big)+\Gamma R_{\ell}\big(\sinh^2{(z_a+z_b-\text{arcsinh}\,\sqrt{\omega})}\big)+R_{\text{si}}\big(\sinh^2{z_b}\big)\Bigg\}
\end{equation}
where the right hand side is minimized over the parameters $z_a$ and $z_b$. Here
\begin{equation}\label{PW_Rsi}
R_{\text{si}}(\omega)=
\begin{cases}
\displaystyle{\frac{1}{2} R_{\text{inf}}(4\omega (1+\omega))} & \text{for }\omega\ge-\frac{1}{2}\\
\displaystyle{-\frac{1}{2} R_{\text{inf}}(4\omega (1+\omega))-\frac{1}{\sqrt{\pi}}\zeta\bigg(\frac{3}{2}\bigg)} & \text{for }\omega\le-\frac{1}{2}
\end{cases}
\end{equation}
with $R_{\text{inf}}(\omega)$ in \eqref{eq:mu fast inf} and $\zeta(z)$ the Riemann zeta function. (reason behind the subscript `si' will be clear shortly). Despite the piecewise nature, the function $R_{\text{si}}(\omega)$ is analytic for all values of $\omega$.
The function $R_{\ell}(\omega(\lambda,\rho_a,\rho_b))=\mu_{\rm fin}(\lambda,\rho_a,\rho_b,\ell-1)$ in \eqref{eq:variational infinite ssep} is the scgf $\mu_{\rm fin}\asymp T^{-1}\log \langle\e^{\lambda Q_T}\rangle$ of $Q_T$ in SSEP on a finite lattice of $(\ell-1)$ sites with unit hopping rate coupled between two unequal reservoirs of density $\rho_a$ and $\rho_b$.

The variational formula \eqref{eq:variational infinite ssep} is intuitively seen from an additivity argument \cite{2021_Derrida_Large}, where the entire infinite line is thought to be composed of three subsystems: a finite SSEP of $(\ell-1)$ sites with hopping rate $\gamma=\Gamma/\sqrt{T}$ coupled with two semi-infinite SSEP with unit hopping rate. The $R_{\text{si}}(\omega(\lambda,\rho_a,\rho_b))=\mu_{\text{si}}(\lambda,\rho_a,\rho_b)$ in \eqref{PW_Rsi} is the scgf of $Q_T$ in a semi-infinite SSEP \cite{2026_Sharma_Large} with unit hopping rate across all bonds. See later discussions. 

Several remarks are in order.

\textit{Remark 1:} 
For $\Gamma\to 0$, the slow region is the bottleneck for particle transport. In this limit, the variational solution \eqref{eq:variational infinite ssep} corresponds to $z_{a(b)}=0$, resulting in $R_{\text{inf}}^{\text{slow}}(\omega,\Gamma,\ell)\sim \Gamma R_{\ell}\big(\omega\big)$, confirming the intuition that the current statistics is governed by the slow region and $\langle\e^{\lambda Q_T}\rangle \asymp\e^{T\gamma R_{\ell}(\omega)}$, where we used $\Gamma=\gamma \sqrt{T}$. (Note the asymptotic dependence on time $T$ compared to \eqref{eq:generating function infinite}.) This large deviation statistics holds true also for $\gamma$ slower than $\sim 1/\sqrt{T}$, and given by the the defect bond effectively coupled between reservoirs of density $\rho_a$ and $\rho_b$.

\textit{Remark 2:} In the other limit $\Gamma \rightarrow \infty$, the variational solution \eqref{eq:variational infinite ssep} corresponds to $z_a=z_b=\frac{1}{2}\text{arcsinh}\sqrt{\omega}$, leading to 
\begin{equation}
R_{\text{inf}}^{\text{slow}}(\omega,\Gamma,\ell)=2R_{\text{si}}\left(\frac{-1+\sqrt{1+\omega}}{2} \right)\qquad \text{for $\Gamma\to \infty$,}
\end{equation}
which reproduces the known expression of $R_{\text{inf}}(\omega)$ in \eqref{eq:mu fast inf} for SSEP without defect bonds. This gives the asymptotics (\ref{eq:generating function infinite},\ref{eq:mu fast inf}), demonstrating that the long-time asymptotics of the current fluctuations, including large deviations, is unaffected by defect bonds unless the jump rate is as slow as $\sim 1/\sqrt{T}$.

\begin{figure}[t]
\centering
\includegraphics[width=\textwidth]{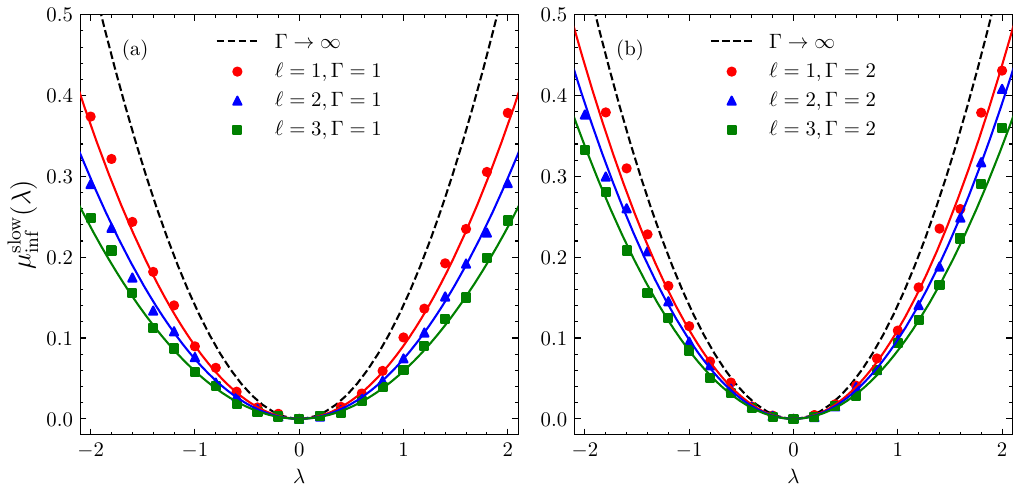}
\caption{\textbf{Scgf for infinite SSEP:} A quantitative comparison of the scgf \eqref{1_slow_inf_scgf} of the time integrated current $Q_T$ across the origin for a half-filled ($\rho_{a(b)}=1/2$) SSEP on an infinite one-dimensional lattice with $\ell$ slow bonds with jump rates $\gamma=\frac{\Gamma}{\sqrt{T}}$ localized around the origin (see Fig.~\ref{fig:inf_with_l_slow_bonds}). Values of the parameters are indicated in the legends. The lines represent theoretical result \eqref{1_slow_inf_scgf}, while the data points represent numerical results from rare-event simulations. Details of the simulation method is discussed in Sec.~\ref{sec:simulation}. The noticeable deviations from the theoretical results near large $\lambda$ is due to finite size effects. The dashed line for $\Gamma\to \infty$ corresponds to the scgf in \eqref{eq:mu fast inf} without slow bonds. Evidently larger number of slow bonds increases the fluctuations in $Q_T$. }
\label{fig:inf_slow_cgf}
\end{figure}

\textit{Remark 3:} Even though the scgf $\mu_{\rm fin}$ for finite SSEP has been formally solved \cite{2012_Gorissen_Current}, to our knowledge, no explicit expression for $R_{\ell}(\omega)$ is known for arbitrary $\ell$. This severely limits us from explicitly solving the variational formula \eqref{eq:variational infinite ssep} for arbitrary number $\ell$ of defect bonds. For the specific cases of $\ell=\{1,2\}$, an explicit formula for the scgf $R_{\ell}(\omega)$ can be obtained from the largest eigenvalue of the corresponding tilted Markov matrix \cite{2004_Derrida_Current}. For the simplest case $\ell=1$, which corresponds to a single bond coupled between two reservoirs, $R_{1}(\omega)=\omega$. The $\ell=2$ case corresponds to a single site coupled between two reservoirs for which
$R_{2}(\omega)=\sqrt{\omega+1}-1$. In these two cases, the variational problem \eqref{eq:variational infinite ssep} can be explicitly solved for the \emph{uniform} initial state ($\rho_{a(b)}=1/2$) yielding a parametric expression,\begin{subequations}\label{1_slow_inf_scgf}
\begin{equation}\label{1_slow_inf_scgf_expr}
R_{\text{inf}}^{\text{slow}}(\omega)=\ell \;\Gamma\sinh^2{\bigg(\frac{4z-\lambda}{2\ell}\bigg)}+\frac{1}{\pi}\int_{-\infty}^\infty\dd k\log{\big(1+\sinh^2{(2z)}\e^{-k^2}\big)}
\end{equation}
for $\ell=\{1,2\}$, where the parameter $z$ is determined in terms of fugacity $\lambda$ using
\begin{equation}\label{1_slow_inf_scgf_param}
\lambda=4z+\ell\; \text{arcsinh}\bigg(\frac{\sinh{(4z)}}{2\pi\Gamma}\int_{-\infty}^\infty\frac{\dd k}{\sinh^2{(2z)}+\e^{k^2}}\bigg).
\end{equation}\end{subequations}
For a large number of slow bonds ($1\ll\ell\ll \sqrt{T}$), using the known explicit result \cite{2004_Bodineau_Current,2004_Derrida_Current} for $R_{\ell}(\omega)\simeq \ell^{-1}(\textrm{arcsinh}\sqrt{\omega})^2$, the variational formula \eqref{eq:variational infinite ssep} gives the scgf. For finite $\ell$, the expression $R_{\ell}(\omega)\simeq \ell\sinh^2(\frac{\textrm{arcsinh}\sqrt{\omega}}{\ell})$ serves as an interpolating formula between small and large $\ell$ regimes, also leading to the same formal expression \eqref{1_slow_inf_scgf}, which provides a reasonable first approximation for arbitrary $\ell\ll \sqrt{T}$.

The exact expression \eqref{1_slow_inf_scgf} for $\ell=\{1,2\}$ is confirmed by numerical simulation using an importance sampling method based on cloning algorithm \cite{2006_Giardinà_Direct, 2007_Lecomte_Numerical, 2019_Perez_Sampling}. A quantitative comparison of the results are shown in Fig. \ref{fig:inf_slow_cgf}. Surprisingly, for $\ell=3$, the formula \eqref{1_slow_inf_scgf} also compares remarkably well, despite not being exact. 

A point to note that the precise position of the slow bonds around the origin is irrelevant at large time $T$. For example, for $\ell=1$, the single defect bond could be between the sites $\{-1,0\}$ or between $\{0,1\}$, without changing the results for the scgf in \eqref{1_slow_inf_scgf}.

\subsection{SSEP on a semi-infinite lattice} 
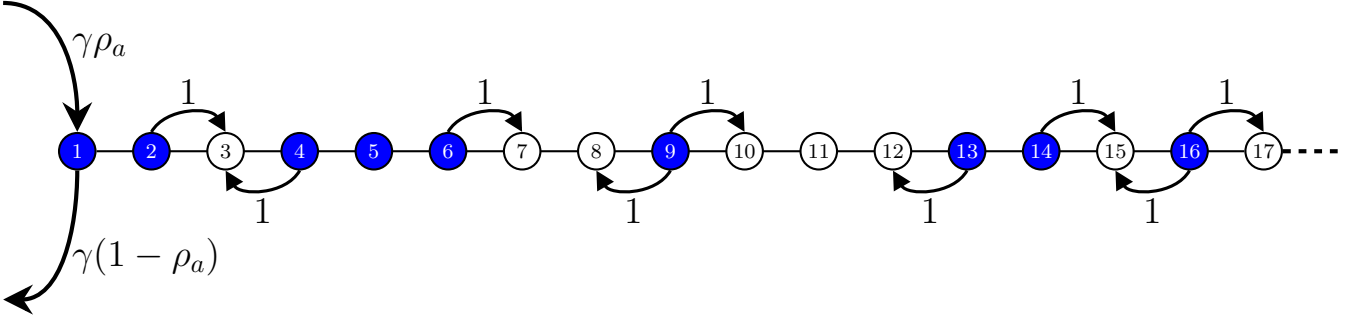
\begin{figure}[t]
\centering
\resizebox{\linewidth}{!}
{
\begin{tikzpicture}
\node [circle, draw=black, thick, fill=blue, inner sep=0pt, minimum size=14pt] (0) at (0,0) {\color{white}\footnotesize1};
\node [circle, draw=black, thick, fill=white, inner sep=0pt, minimum size=14pt] (16) at (16,0) {\footnotesize17};

\draw[thick] (0) -- (16);
\draw[ultra thick, dashed] (16) -- (17,0);

\node [circle, draw=black, thick, fill=blue, inner sep=0pt, minimum size=14pt] (1) at (1,0) {\color{white}\footnotesize2};
\node [circle, draw=black, thick, fill=white, inner sep=0pt, minimum size=14pt] (2) at (2,0) {\footnotesize3};
\node [circle, draw=black, thick, fill=blue, inner sep=0pt, minimum size=14pt] (3) at (3,0) {\color{white}\footnotesize4};
\node [circle, draw=black, thick, fill=blue, inner sep=0pt, minimum size=14pt] (4) at (4,0) {\color{white}\footnotesize5};
\node [circle, draw=black, thick, fill=blue, inner sep=0pt, minimum size=14pt] (5) at (5,0) {\color{white}\footnotesize6};
\node [circle, draw=black, thick, fill=white, inner sep=0pt, minimum size=14pt] (6) at (6,0) {\footnotesize7};
\node [circle, draw=black, thick, fill=white, inner sep=0pt, minimum size=14pt] (7) at (7,0) {\footnotesize8};
\node [circle, draw=black, thick, fill=blue, inner sep=0pt, minimum size=14pt] (8) at (8,0) {\color{white}\footnotesize9};
\node [circle, draw=black, thick, fill=white, inner sep=0pt, minimum size=14pt] (9) at (9,0) {\footnotesize10};
\node [circle, draw=black, thick, fill=white, inner sep=0pt, minimum size=14pt] (10) at (10,0) {\footnotesize11};
\node [circle, draw=black, thick, fill=white, inner sep=0pt, minimum size=14pt] (11) at (11,0) {\footnotesize12};
\node [circle, draw=black, thick, fill=blue, inner sep=0pt, minimum size=14pt] (12) at (12,0) {\color{white}\footnotesize13};
\node [circle, draw=black, thick, fill=blue, inner sep=0pt, minimum size=14pt] (13) at (13,0) {\color{white}\footnotesize14};
\node [circle, draw=black, thick, fill=white, inner sep=0pt, minimum size=14pt] (14) at (14,0) {\footnotesize15};
\node [circle, draw=black, thick, fill=blue, inner sep=0pt, minimum size=14pt] (15) at (15,0) {\color{white}\footnotesize16};

\draw [-{Stealth[width=4.0mm, length=4.0mm]}, ultra thick] (-1,2) to [out=0,in=90] node [midway, right]{\Large$\gamma\rho_a$} (0.north);
\draw [{Stealth[width=4.0mm, length=4.0mm]}-, ultra thick] (-1,-2) to [out=0,in=-90] node [midway, right]{\Large$\gamma(1-\rho_a)$} (0.south);
\draw [-{Triangle[width=2.5mm, length=2.5mm]}, very thick] (1.north) to [out=60,in=120] node [midway, above]{\Large$1$} (2.north);
\draw [-{Triangle[width=2.5mm, length=2.5mm]}, very thick] (3.south) to [out=-120,in=-60] node [midway, below]{\Large$1$} (2.south);
\draw [-{Triangle[width=2.5mm, length=2.5mm]}, very thick] (5.north) to [out=60,in=120] node [midway, above]{\Large$1$} (6.north);
\draw [-{Triangle[width=2.5mm, length=2.5mm]}, very thick] (8.north) to [out=60,in=120] node [midway, above]{\Large$1$} (9.north);
\draw [-{Triangle[width=2.5mm, length=2.5mm]}, very thick] (8.south) to [out=-120,in=-60] node [midway, below]{\Large$1$} (7.south);
\draw [-{Triangle[width=2.5mm, length=2.5mm]}, very thick] (12.south) to [out=-120,in=-60] node [midway, below]{\Large$1$} (11.south);
\draw [-{Triangle[width=2.5mm, length=2.5mm]}, very thick] (13.north) to [out=60,in=120] node [midway, above]{\Large$1$} (14.north);
\draw [-{Triangle[width=2.5mm, length=2.5mm]}, very thick] (15.south) to [out=-120,in=-60] node [midway, below]{\Large$1$} (14.south);
\draw [-{Triangle[width=2.5mm, length=2.5mm]}, very thick] (15.north) to [out=60,in=120] node [midway, above]{\Large$1$} (16.north);
\end{tikzpicture}
}
\caption{\textbf{Semi-infinite SSEP:} SSEP on a positive integer lattice coupled to a reservoir of density $\rho_a$ at the leftmost site through a bond whose jump rate is controlled by a parameter $\gamma$. In the bulk, jump rates are uniformly set to value one.}
\label{fig:semi_inf_with_slow_boundary_coupling}
\end{figure}

In this example the SSEP is defined on a one-dimensional positive integer lattice with the sites labeled by $i=1,2,3,\cdots$. At the leftmost site ($i=1$), particles are injected at a rate of $\gamma\rho_a$ following exclusion, and removed at a rate of $\gamma(1-\rho_a)$. This models coupling to a particle reservoir of density $\rho_a$ with $\gamma$ dictating the time scale. In the bulk, particles hop between adjacent sites at a uniform rate of one, following exclusion. A schematic is shown in Fig.~\ref{fig:semi_inf_with_slow_boundary_coupling}.

The lattice is initially populated according to a Bernoulli product measure with uniform density $\rho_b$. With time, particles are exchanged between the lattice and the reservoir. We denote by $Q_T$ the net flow of particles from the reservoir to the lattice in time $T$. 

\subsubsection{Fast coupling regime}

For $\gamma=1$, the long-time statistics of $Q_T$ follows a similar statistics as in the infinite-line example, where the generating function 
\begin{equation}\label{eq:gen fnc si}
\big<\e^{\lambda Q_T}\big>\asymp\e^{\sqrt{T}\mu_{\text{si}}(\lambda,\rho_a,\rho_b)},
\end{equation}
with the scgf $\mu_{\text{si}}(\lambda,\rho_a,\rho_b)=R_{\text{si}}(\omega(\lambda,\rho_a,\rho_b))$ in \eqref{PW_Rsi} obtained recently \cite{2026_Sharma_Large} using MFT.

\begin{figure}[t]
\centering
\includegraphics[width=0.49\linewidth]{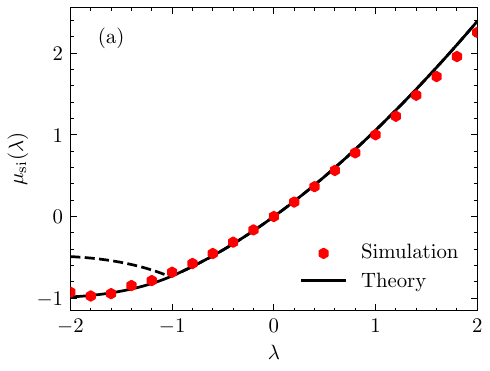}
\includegraphics[width=0.495\textwidth]{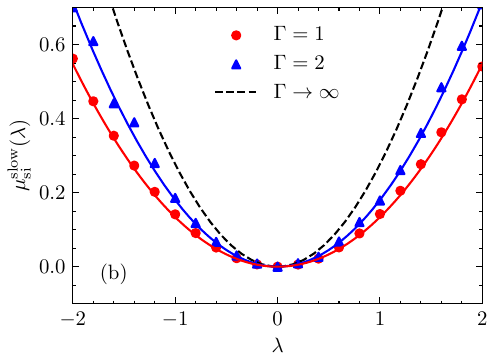}
\caption{\textbf{Scgf for semi-infinite SSEP:} (a) The scgf of current for the semi-infinite SSEP in the fast coupling regime ($\gamma=1$) for densities $(\rho_a,\rho_b)\equiv(0.9,0.1)$. The solid black line represents the theoretical result $\mu_{\text{si}}(\lambda,\rho_a,\rho_b)=R_{\text{si}}(\omega(\lambda,\rho_a,\rho_b))$ with \eqref{PW_Rsi}, while the data points are obtained by a cloning algorithm with $T=500$ and $N_c=50,000$. The dashed line represents the function $\frac{1}{2}R_{\text{inf}}(4\omega (1+\omega))$ for parameter regimes of $\lambda$ where $\omega<-1/2$, asserting the piece-wise function in \eqref{PW_Rsi}. (b) Comparison of the scgf \eqref{eq:mu si half} for half-filled SSEP in the slow coupling regime. The dashed curve for $\Gamma\to \infty$ represents the scgf \eqref{PW_Rsi} in the fast-coupling regime, while the solid lines correspond to $\gamma=\frac{\Gamma}{\sqrt{T}}$ with values of $\Gamma$ indicated in the plot legends. The data points represent numerical results from rare-event simulations using cloning algorithm.}
\label{fig:semi_inf_slow_cgf}
\end{figure}

A variational construction similar to \eqref{eq:variational infinite ssep} offers a far simpler derivation of \eqref{PW_Rsi} compared to the involved methods in \cite{2026_Sharma_Large,2024_Grabsch}. Following the additivity argument in \cite{2021_Derrida_Large} the infinite line SSEP can be thought of as composed of two semi-infinite line SSEPs with the current large deviation function given by addition of corresponding large deviation functions of two semi-infinite SSEPs optimized over the density at the contact point. An equivalent statement in terms of corresponding scgf,
\begin{equation}\label{eq:variational semi-infinite ssep}
\mu_{\text{inf}}(\lambda,\rho_a,\rho_b)=\min_{\lambda_0}\max_{\rho_0}\{\mu_{\text{si}}(\lambda_0,\rho_a,\rho_0)+\mu_{\text{si}}(\lambda-\lambda_0,\rho_0,\rho_b)\}
\end{equation}
Considering that the scgfs are functions of $\omega$ and an identity \eqref{eq: variational_id}, the above variational problem reduces to 
\begin{equation}\label{eq:variational infinite ssep 2}
R_{\text{inf}}(\omega)=\min_{z}\{R_{\text{si}}\big(\sinh^2{z}\big)+R_{\text{si}}\big(\sinh^2{(z\pm u)}\big)\}
\end{equation}
with $\omega(\lambda,\rho_a,\rho_b)=\sinh^2 u$. The optimization is for $z=\frac{u}{2}$ leading to 
\begin{equation}\label{eq:si inf relation}
R_{\text{inf}}(\omega)=2R_{\text{si}}\left(\frac{-1+\sqrt{1+\omega}}{2}\right)
\end{equation}
The relation is valid for $\omega\ge -1$ where the argument of $R_{\rm si}$ is real. This  reproduces the formula \eqref{PW_Rsi} expressing $R_{\text{si}}(\omega)$ in terms of $R_{\text{inf}}(\omega)$ for $\omega\ge -\frac{1}{2}$. The known expression \eqref{eq:mu fast inf} for $R_{\text{inf}}(\omega)$ solves for $R_{\text{si}}(\omega)$. The complete formula as a piece-wise function \eqref{PW_Rsi} is constructed \cite{2026_Sharma_Large} by analytical continuation demanding convexity of the scgf and confirmed in numerical simulation shown in Fig.~\ref{fig:semi_inf_slow_cgf}a.

\subsubsection{Slow coupling regime}

Similar to the infinite-line case, the fast coupling result remains unchanged for $\gamma$ larger than $\sim 1/\sqrt{T}$. In the marginal case, $\gamma=\frac{\Gamma}{\sqrt{T}}$ with $\Gamma$ a positive valued parameter, we write the corresponding scgf $\mu_{\text{si}}^{\rm slow}(\lambda,\rho_a,\rho_b)=R_{\text{si}}^{\text{slow}}(\omega(\lambda,\rho_a,\rho_b),\Gamma)$ in a variational formula:
\begin{equation}\label{eq:var si}
R_{\text{si}}^{\text{slow}}(\omega,\Gamma)=\min_{z}\Big\{\Gamma\sinh^2{(z-\text{arcsinh}\,\sqrt{\omega})}+R_{\text{si}}\big(\sinh^2{z}\big)\Big\}
\end{equation}
with $R_{\text{si}}(\omega)$ defined in \eqref{PW_Rsi}.

For $\Gamma\to 0$, the variational solution corresponds to $z=0$, resulting $R_{\text{si}}^{\text{slow}}(\omega,\Gamma)\simeq \Gamma \omega$ which is proportional to $R_1(\omega)=\omega$ defined earlier in \eqref{1_slow_inf_scgf}. Thus, $\langle\mathrm{e}^{\lambda Q_T} \rangle \asymp\mathrm{e}^{T\gamma\omega}$, showing that the current statistics is essentially governed by transport across the single slow bond coupling the reservoir and the system. The $\Gamma\to \infty$ limit corresponds to the fast coupling regime with the asymptotic \eqref{eq:gen fnc si} and the scgf $\mu_{\text{si}}(\lambda,\rho_a,\rho_b)=R_{\text{si}}(\omega(\lambda,\rho_a,\rho_b))$. This is seen from the variational solution $z=\text{arcsinh}\,\sqrt{\omega}$ of \eqref{eq:var si} in the $\Gamma\to \infty$ limit.

For the half-filled case ($\rho_{a(b)}=1/2$) the expression particularly simplifies leading to the scgf 
\begin{equation}\label{eq:mu si half}
\mu_{\text{si}}^{\text{slow}}\left(\lambda,\frac{1}{2},\frac{1}{2}\right)=\min_z\Bigg\{\Gamma\sinh^2{\bigg(z-\frac{\lambda}{2}\bigg)}+\frac{1}{\pi}\int_{-\infty}^\infty\dd k\log{\big[1+\sinh^2{(2z)}\,\e^{-k^2}\big]}\Bigg\}
\end{equation}
This expression is confirmed using rare event simulations following cloning algorithm and the results are shown in Fig.~\ref{fig:semi_inf_slow_cgf}b.

\subsection{SSEP on a finite lattice coupled to boundary reservoirs \label{sec:finite ssep}}

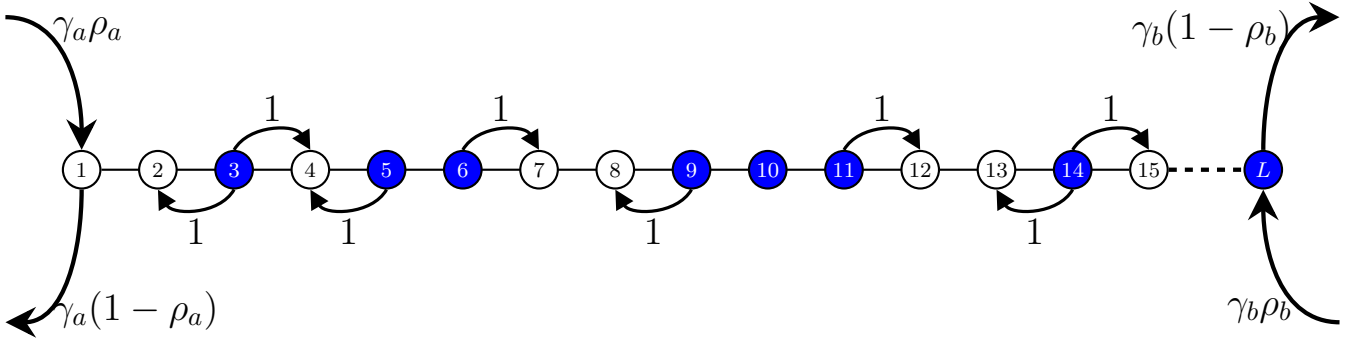
\begin{figure}[t]
\centering
\resizebox{\linewidth}{!}
{
\begin{tikzpicture}
\node [circle, draw=black, thick, fill=white, inner sep=0pt, minimum size=14pt] (0) at (0,0) {\footnotesize1};
\node [circle, draw=black, thick, fill=white, inner sep=0pt, minimum size=14pt] (14) at (14,0) {\footnotesize15};
\node [circle, draw=black, thick, fill=blue, inner sep=0pt, minimum size=14pt] (15) at (15.5,0) {\color{white}\footnotesize$L$};

\draw[thick] (0) -- (14);
\draw[ultra thick, dashed] (14) -- (15);

\node [circle, draw=black, thick, fill=white, inner sep=0pt, minimum size=14pt] (1) at (1,0) {\footnotesize2};
\node [circle, draw=black, thick, fill=blue, inner sep=0pt, minimum size=14pt] (2) at (2,0) {\color{white}
\footnotesize3};
\node [circle, draw=black, thick, fill=white, inner sep=0pt, minimum size=14pt] (3) at (3,0) {\footnotesize4};
\node [circle, draw=black, thick, fill=blue, inner sep=0pt, minimum size=14pt] (4) at (4,0) {\color{white}\footnotesize5};
\node [circle, draw=black, thick, fill=blue, inner sep=0pt, minimum size=14pt] (5) at (5,0) {\color{white}\footnotesize6};
\node [circle, draw=black, thick, fill=white, inner sep=0pt, minimum size=14pt] (6) at (6,0) {\footnotesize7};
\node [circle, draw=black, thick, fill=white, inner sep=0pt, minimum size=14pt] (7) at (7,0) {\footnotesize8};
\node [circle, draw=black, thick, fill=blue, inner sep=0pt, minimum size=14pt] (8) at (8,0) {\color{white}\footnotesize9};
\node [circle, draw=black, thick, fill=blue, inner sep=0pt, minimum size=14pt] (9) at (9,0) {\color{white}\footnotesize10};
\node [circle, draw=black, thick, fill=blue, inner sep=0pt, minimum size=14pt] (10) at (10,0) {\color{white}\footnotesize11};
\node [circle, draw=black, thick, fill=white, inner sep=0pt, minimum size=14pt] (11) at (11,0) {\footnotesize12};
\node [circle, draw=black, thick, fill=white, inner sep=0pt, minimum size=14pt] (12) at (12,0) {\footnotesize13};
\node [circle, draw=black, thick, fill=blue, inner sep=0pt, minimum size=14pt] (13) at (13,0) {\color{white}\footnotesize14};

\draw [-{Stealth[width=4.0mm, length=4.0mm]}, ultra thick] (-1,2) to [out=0,in=90] node [near start, right]{\Large$\gamma_a\rho_a$} (0.north);
\draw [{Stealth[width=4.0mm, length=4.0mm]}-, ultra thick] (-1,-2) to [out=0,in=-90] node [near start, right]{\Large$\gamma_a(1-\rho_a)$} (0.south);

\draw [-{Triangle[width=2.5mm, length=2.5mm]}, very thick] (2.south) to [out=-120,in=-60] node [midway, below]{\Large$1$} (1.south);
\draw [-{Triangle[width=2.5mm, length=2.5mm]}, very thick] (2.north) to [out=60,in=120] node [midway, above]{\Large$1$} (3.north);
\draw [-{Triangle[width=2.5mm, length=2.5mm]}, very thick] (4.south) to [out=-120,in=-60] node [midway, below]{\Large$1$} (3.south);
\draw [-{Triangle[width=2.5mm, length=2.5mm]}, very thick] (5.north) to [out=60,in=120] node [midway, above]{\Large$1$} (6.north);
\draw [-{Triangle[width=2.5mm, length=2.5mm]}, very thick] (8.south) to [out=-120,in=-60] node [midway, below]{\Large$1$} (7.south);
\draw [-{Triangle[width=2.5mm, length=2.5mm]}, very thick] (10.north) to [out=60,in=120] node [midway, above]{\Large$1$} (11.north);
\draw [-{Triangle[width=2.5mm, length=2.5mm]}, very thick] (13.south) to [out=-120,in=-60] node [midway, below]{\Large$1$} (12.south);
\draw [-{Triangle[width=2.5mm, length=2.5mm]}, very thick] (13.north) to [out=60,in=120] node [midway, above]{\Large$1$} (14.north);

\draw [{Stealth[width=4.0mm, length=4.0mm]}-, ultra thick] (16.5,2) to [out=180,in=90] node [near start, left]{\Large$\gamma_b(1-\rho_b)$} (15.north);
\draw [-{Stealth[width=4.0mm, length=4.0mm]}, ultra thick] (16.5,-2) to [out=180,in=-90] node [near start, left]{\Large$\gamma_b\rho_b$} (15.south);
\end{tikzpicture}
}
\caption{\textbf{Finite SSEP:} SSEP on a finite lattice coupled to two boundary reservoirs of densities $\rho_a$ and $\rho_b$ with slow bonds.}
\label{fig:fin_with_slow_boundary_coupling}
\end{figure}

\begin{figure}[t]
\centering
\includegraphics[width=0.6\textwidth]{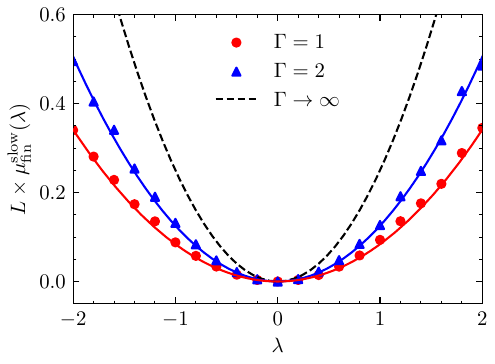}
\caption{\textbf{Scgf for finite SSEP:} Comparison of the scgf of the empirical current $Q_T$ transferred across a large system of length $L=32$ between two boundary reservoirs of density $\rho_{a(b)}=\frac{1}{2}$. The jump rates across the system-reservoir bonds is parametrized by $\gamma=\frac{\Gamma}{L}$ with values of $\Gamma$ indicated in the plot legends. The solid lines are the theoretical results \eqref{eq:R fin half} for different $\Gamma$, and the data points are the corresponding results from rare-event simulations using cloning algorithm. The $\Gamma\to \infty$ corresponds to the fast coupling result $\left(\textrm{arcsinh}\sqrt{\omega(\lambda,\rho_a,\rho_b)}\right)^2$.}
\label{fig: finite_slow_cgf}
\end{figure}

This third example is about SSEP on a finite integer lattice of $L$ sites indexed by $i=1,2,\cdots,L-1,L$. Across the bonds inside the bulk,
particles hop with uniform unit rate following simple exclusion. At the leftmost site ($i=1$), a particle is added and removed at a rate of $\gamma\rho_a$ and $\gamma(1-\rho_a)$, respectively, while at the rightmost site ($i=L$), a particle is added and removed at a rate of $\gamma\rho_b$ and $\gamma(1-\rho_b)$, respectively. This choice of rates corresponds to the left and right boundary sites coupled to particle reservoirs of densities $\rho_a$ and $\rho_b$, respectively. The positive valued parameter $\gamma$ controls the time scale of coupling with the reservoirs.

At long-times the system eventually reaches a NESS characterized by a constant average particle flux between the two reservoirs. 
In this steady state the net number of particles $Q_T$ transferred over a long period $T$ from the left reservoir to the right reservoir, follows a large deviation statistics, where the generating function
\begin{equation}
\big<\e^{\lambda Q_T}\big>\sim\e^{T\,\mu_{\text{fin}}(\lambda,\rho_a,\rho_b,L)}
\end{equation}
for large $T$, with the corresponding scgf $\mu_{\text{fin}}$ depending on $(\lambda,\rho_a,\rho_b)$ through a single function $\omega(\lambda,\rho_a,\rho_b)$ defined in \eqref{omega_param_defn}. Unlike the corresponding problems on the infinite and semi-infinite lines, the statistics does not depend on the details of the initial state.

In general, the scgf depends on the coupling parameter $\gamma$. For $\gamma=1$ and large $L$, the scgf $\mu_{\text{fin}}(\lambda,\rho_a,\rho_b,L)= L^{-1}\left(\textrm{arcsinh}\sqrt{\omega(\lambda,\rho_a,\rho_b)}\right)^2$ was obtained in \cite{2004_Bodineau_Current} using additivity principle and in \cite{2004_Derrida_Current} using a tilted matrix formalism. For arbitrary $\gamma$, two distinct scenarios emerge: (a) for $\gamma$ larger than $\sim 1/L$, the current statistics is governed by the bulk transport and described by the same asymptotic for $\gamma=1$, (b) for $\gamma $ smaller than $\sim 1/L$, the two defect bonds coupling the system with reservoirs act as the bottleneck and the current statistics is effectively given by a single site coupled between two reservoirs, leading to $\mu_{\text{fin}}(\lambda,\rho_a,\rho_b,L)= \gamma R_2(\omega(\lambda,\rho_a,\rho_b))$ independent of $L$ with $R_2(\omega)=\sqrt{\omega+1}-1$.

Both scenarios can be obtained from different limits of the marginal case $\gamma=\frac{\Gamma}{L}$ with positive valued parameter $\Gamma$. Corresponding scgf was computed \cite{2021_Derrida_Large} following an additivity argument where the system is assumed to be composed of three sub-components: the bulk and the two slow bonds coupling the bulk with the reservoirs. For large $L$, the result is expressed in a variational formula for scgf $\mu_{\text{fin}}^{\text{slow}}(\lambda,\rho_a,\rho_b,L)=R_{L}^{\text{slow}}(\omega(\lambda,\rho_a,\rho_b))$ with
\begin{equation}\label{eq:var sol fin}
R_{L}^{\text{slow}}(\omega)\simeq \frac{1}{L}\min_{z_a,z_b}\Big\{\Gamma\sinh^2{z_a}+(z_a+z_b-\text{arcsinh}\,\sqrt{\omega})^2+\Gamma\sinh^2{z_b}\Big\}.
\end{equation}
We have confirmed the result~\eqref{eq:var sol fin} using numerical simulation following cloning algorithm and shown in Fig.~\ref{fig: finite_slow_cgf}. 

\textit{Remark 1:} In the extreme slow coupling regime $\Gamma\to 0$, variational solution of \eqref{eq:var sol fin} corresponds to $z_{a}+z_b=\text{arcsinh}\,\sqrt{\omega}$, leading to 
\begin{equation}
R_L^{\text{slow}}(\omega)\simeq \frac{\Gamma}{L}\min_z\left\{\sinh^2 z + \sinh^2(\text{arcsinh}\,\sqrt{\omega}-z)\right\}=\frac{\Gamma}{L}\left\{-1+\sqrt{1+\omega}\right\},
\end{equation}
confirming the result discussed above in scenario (b) with $\langle\e^{\lambda Q_T}\rangle \asymp\e^{T\gamma R_2(\omega)}$. In the opposite limit $\Gamma\to\infty$, the variational solution \eqref{eq:var sol fin} corresponds to $z_{a(b)}=0$, leading to the fast coupling result $R_L(\omega)\simeq L^{-1}(\textrm{arcsinh}\sqrt{\omega})^2$ discussed in scenario (a).

\textit{Remark 2:} For the half-filled case ($\rho_{a(b)}=1/2$) the expression \eqref{eq:var sol fin} simplifies to a parametric expression 
\begin{equation}\label{eq:R fin half}
R_L^{\text{slow}}(\omega)=\frac{1}{L}\left\{2\,\Gamma\sinh^2{z}+\bigg(2z-\frac{\lambda}{2}\bigg)^2\right\}\quad \text{with }\lambda=4z+\Gamma\sinh{(2z)}
\end{equation}
used for the plots in Fig.~\ref{fig: finite_slow_cgf}.

\section{Macroscopic fluctuation theory for SSEP with slow bonds on an infinite lattice \label{sec:inf SSEP}}
Our results in Sec.~\ref{sec: SOR} are obtained using the hydrodynamics approach of MFT suitably modified to incorporate localized defect bonds. In this Section, we present a bottom-up construction of the theory for SSEP on an infinite line and a derivation of the expression \eqref{eq:variational infinite ssep}. Our construction is based on a similar route presented in our earlier works \cite{2026_Saha,2024_Saha_Large} about lattice and off-lattice dynamics.

On the infinite one-dimensional lattice with sites indexed by integer $i$, we specify a configuration by the binary occupation variables $\boldsymbol{n}(\tau)\equiv\{n_i(\tau)\}$ with $n_i(\tau)=0$ or $1$ depending on whether the $i$-th site at time $\tau$ is vacant or filled. 
We denote $Y_i(\tau)$ as the net number of particle-jumps from site $i$ to $i+1$ in the infinitesimal time window between $\tau$ and $\tau+\dd\tau$.

\begin{figure}[t]
\centering
\resizebox{\linewidth}{!}
{
\begin{tikzpicture}
\node [circle, draw=black, thick, fill=white, inner sep=0pt, minimum size=14pt] (1) at (1,0) {\footnotesize-7};
\node [circle, draw=black, thick, fill=blue, inner sep=0pt, minimum size=14pt] (8) at (8,0) {\color{white}\footnotesize0};
\node [circle, draw=black, thick, fill=white, inner sep=0pt, minimum size=14pt] (9) at (9,0) {\footnotesize1};
\node [circle, draw=black, thick, fill=blue, inner sep=0pt, minimum size=14pt] (16) at (16,0) {\color{white}\footnotesize8};

\draw[ultra thick, dashed] (1) -- (0,0);
\draw[ultra thick] (1) -- (8);
\draw[thick, decorate, decoration={zigzag, segment length=2.2pt, amplitude=1.1pt}] (8) -- (9);
\draw[thick] (9) -- (16);
\draw[ultra thick, dashed] (16) -- (17,0);

\node [circle, draw=black, thick, fill=blue, inner sep=0pt, minimum size=14pt] (2) at (2,0) {\color{white}\footnotesize-6};
\node [circle, draw=black, thick, fill=white, inner sep=0pt, minimum size=14pt] (3) at (3,0) {\footnotesize-5};
\node [circle, draw=black, thick, fill=blue, inner sep=0pt, minimum size=14pt] (4) at (4,0) {\color{white}\footnotesize-4};
\node [circle, draw=black, thick, fill=blue, inner sep=0pt, minimum size=14pt] (5) at (5,0) {\color{white}\footnotesize-3};
\node [circle, draw=black, thick, fill=blue, inner sep=0pt, minimum size=14pt] (6) at (6,0) {\color{white}\footnotesize-2};
\node [circle, draw=black, thick, fill=white, inner sep=0pt, minimum size=14pt] (7) at (7,0) {\footnotesize-1};

\node [circle, draw=black, thick, fill=blue, inner sep=0pt, minimum size=14pt] (10) at (10,0) {\color{white}\footnotesize2};
\node [circle, draw=black, thick, fill=white, inner sep=0pt, minimum size=14pt] (11) at (11,0) {\footnotesize3};
\node [circle, draw=black, thick, fill=blue, inner sep=0pt, minimum size=14pt] (12) at (12,0) {\color{white}\footnotesize4};
\node [circle, draw=black, thick, fill=blue, inner sep=0pt, minimum size=14pt] (13) at (13,0) {\color{white}\footnotesize5};
\node [circle, draw=black, thick, fill=white, inner sep=0pt, minimum size=14pt] (14) at (14,0) {\footnotesize6};
\node [circle, draw=black, thick, fill=white, inner sep=0pt, minimum size=14pt] (15) at (15,0) {\footnotesize7};

\draw [-{Triangle[width=2.5mm, length=2.5mm]}, very thick] (2.south) to [out=-120,in=-60] node [midway, below]{\Large$\alpha$} (1.south);
\draw [-{Triangle[width=3.0mm, length=3.0mm]}, very thick] (2.north) to [out=60,in=120] node [midway, above]{\Large$\alpha$} (3.north);
\draw [-{Triangle[width=3.0mm, length=3.0mm]}, very thick] (4.south) to [out=-120,in=-60] node [midway, below]{\Large$\alpha$} (3.south);
\draw [-{Triangle[width=3.0mm, length=3.0mm]}, very thick] (6.north) to [out=60,in=120] node [midway, above]{\Large$\alpha$} (7.north);
\draw [-{Triangle[width=3.0mm, length=3.0mm]}, very thick] (8.south) to [out=-120,in=-60] node [midway, below]{\Large$\alpha$} (7.south);

\draw [-{Triangle[width=3.0mm, length=3.0mm]}, very thick] (8.north) to [out=60,in=120] node [midway, above]{\Large$\gamma$} (9.north);

\draw [-{Triangle[width=3.0mm, length=3.0mm]}, ultra thick] (10.south) to [out=-120,in=-60] node [midway, below]{\Large$1$} (9.south);
\draw [-{Triangle[width=3.0mm, length=3.0mm]}, very thick] (10.north) to [out=60,in=120] node [midway, above]{\Large$1$} (11.north);
\draw [-{Triangle[width=3.0mm, length=3.0mm]}, ultra thick] (12.south) to [out=-120,in=-60] node [midway, below]{\Large$1$} (11.south);
\draw [-{Triangle[width=3.0mm, length=3.0mm]}, very thick] (13.north) to [out=60,in=120] node [midway, above]{\Large$1$} (14.north);
\draw [-{Triangle[width=3.0mm, length=3.0mm]}, very thick] (16.south) to [out=-120,in=-60] node [midway, below]{\Large$1$} (15.south);
\end{tikzpicture}
}
\caption{Symmetric jump rates in the SSEP on an infinite lattice. Across the defect bond connecting sites $i=0$ and $i=1$, the hopping rate is $\gamma$, while across the bonds to the left (right) of the defect, the hopping rates are uniformly set as $\alpha$ ($1$). Different bond-styles are used for indicating the difference in the jump rates.}\label{fig:inf_ssep_derive}
\end{figure}
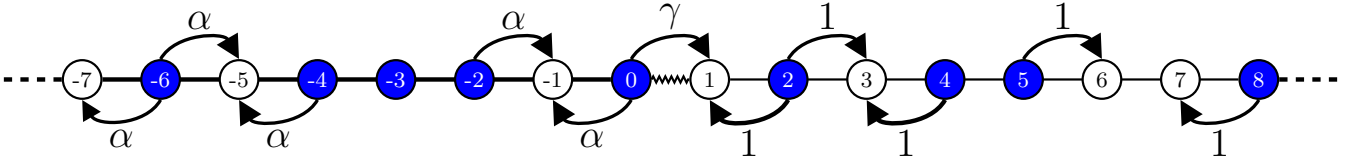

The empirical current across the bond between $i$-th and $(i+1)$-th sites $Q_i(T)=\sum_{k=0}^{M-1}Y_i(k\,\dd \tau)$ where we have divided the time period $[0,T]$ into $M$ infinitesimal segments each of duration $\dd \tau$. The $Y_i(\tau)$ and hence $Q_i(T)$ fluctuate, and the latter statistics is characterized by the generating function 
\begin{equation}\label{eq:bare G inf}
G_T\left(\boldsymbol{\lambda}; \boldsymbol{n}(0),\boldsymbol{n}(T)\right)=\left\langle\e^{\sum_i\lambda_iQ_i(T)}\big\vert\boldsymbol{n}(0),\boldsymbol{n}(T)\right\rangle
\end{equation}
with the fugacity parameter $\boldsymbol{\lambda}\equiv \{\lambda_i\}$ and the angular brackets denoting average over evolutions between the initial and final configurations, $\boldsymbol{n}(0)$ and $\boldsymbol{n}(T)$, respectively. This average can be written as a path integral 
\begin{align}\label{gen_fun_inf_expr1}
G_T&=\int_{\boldsymbol{n}(0)}^{\boldsymbol{n}(T)}\left[\mathcal{D}n\right]\Bigg<\e^{\sum_{k=0}^{M-1}\sum_{i}\lambda_iY_i(k\,\dd \tau)}\prod_{k=0}^{M-1}\,\prod_{i}\Big(\delta_{n_i((k+1)\dd \tau)-n_i(k\,\dd \tau),Y_{i-1}(k\,\dd \tau)-Y_i(k\,\dd \tau)}\Big)\Bigg>_{\boldsymbol{Y}} 
\end{align}
where the path integral measure
\begin{equation}\label{eq:path integral measure}
\int\left[\mathcal{D}n\right]\equiv\prod_{k=1}^{M-1}\prod_i\sum_{n_i(k\,\dd \tau)=0,1}\, ,
\end{equation}
the Kronecker-delta $\delta_{a,b}$ is for the particle-number conservation
\begin{equation}\label{site_bond_conservation}
n_i(\tau+\dd \tau)-n_i(\tau)=Y_{i-1}(\tau)-Y_i(\tau),
\end{equation}
and $\langle\cdots\rangle_{\mathbf{Y}}$ denotes average over histories of $\{Y_i(\tau)\}$.

Considering infinitesimal $\dd \tau$, $Y_i(\tau)$ takes three possible values,
\begin{subnumcases}
{\label{lft_semi_inf}Y_i(\tau)=}
1 & with prob. $p_i\,n_i(\tau)\,\big(1-n_{i+1}(\tau)\big)\,\dd \tau$\\
-1 & with prob. $p_i\,n_{i+1}(\tau)\,\big(1-n_i(\tau)\big)\,\dd \tau$\\
0 & with prob. $1-p_i\,\big\{n_i(\tau)\,\big(1-n_{i+1}(\tau)\big)+n_{i+1}(\tau)\,\big(1-n_i(\tau)\big)\big\}\,\dd \tau$
\end{subnumcases}
where we choose uniform jump rates $p_i=\alpha$ for bonds left of the $0$-th site ($i<0$), $p_0=\gamma$ for the defect bond between sites $0$ and $1$, and $p_i=1$ for the rest ($i\ge1$). See Fig. \ref{fig:inf_ssep_derive} for a schematic of the jump rates.

For computing the average in \eqref{gen_fun_inf_expr1} we use $\delta_{a,b}=(2\pi\mathrm{i})^{-1}\int_{-\pi\mathrm{i}}^{\pi\mathrm{i}}\dd z\,\e^{-z(a-b)}$ and introduce a conjugate variable $\widehat{n}_i(\tau)$, writing
\begin{align}\label{eq:nhat intro}
G_T=\int_{\boldsymbol{n}(0)}^{\boldsymbol{n}(T)}\left[\mathcal{D}n\right]\left[\mathcal{D}\widehat{n}\right]&\exp{\bigg\{-\sum_{k=0}^{M-1}\,\sum_{i}\widehat{n}_i(k\,\dd \tau)\,\big(n_i((k+1)\dd \tau)-n_i(k\,\dd \tau)\big)\bigg\}}\nonumber\\
&\quad\Bigg<\exp{\sum_{k=0}^{M-1}\,\sum_{i}\bigg(\widehat{n}_{i+1}(k\,\dd \tau)-\widehat{n}_i(k\,\dd \tau)+\lambda_i\bigg)\,Y_i(k\,\dd \tau)}\Bigg>_{\boldsymbol{Y}}
\end{align}
where we used $\sum_{i}\widehat{n}_i\,(Y_{i-1}-Y_i)=\sum_{i}(\widehat{n}_{i+1}-\widehat{n}_i)\,Y_i$. The path-integral measure $[\mathcal{D}\widehat{n}]$ is defined analogous to \eqref{eq:path integral measure}.

Averaging over $\mathbf{Y}$ with its distribution \eqref{lft_semi_inf} we write
\begin{equation} \label{gen_fun_inf_expr3}
G_T=\int_{\boldsymbol{n}(0)}^{\boldsymbol{n}(T)}\left[\mathcal{D}n\right]\left[\mathcal{D}\widehat{n}\right]\e^{\mathcal{K}+\mathcal{H}}
\end{equation}
where the terms in the exponential,
\begin{subequations}\label{gen_fun_inf_expr3 v1}\begin{align}
\mathcal{K}&=\sum_{i}\Bigg\{n_i(0)\,\widehat{n}_i(0)-n_i(T)\,\widehat{n}_i(T)+\int_0^T\dd \tau\,n_i(\tau)\,\frac{\dd \widehat{n}_i(\tau)}{\dd \tau}\Bigg\}\\
\mathcal{H}&=\sum_{i}\,p_i\,\int_0^T\dd \tau\,\omega\big(\widehat{n}_{i+1}(\tau)-\widehat{n}_i(\tau)+\lambda_i,n_i(\tau),n_{i+1}(\tau)\big)
\end{align}
\end{subequations}
with the $\omega$ defined in \eqref{omega_param_defn}. This path integral formulation is at the microscopic scale and is complimentary to the Doi-Peliti formulation (see \cite{2006_Andreanov_Exact,2007_Lefevre_Dynamics} and references therein).

\paragraph*{Current across the origin:} Our results in Sec.~\ref{sec: SOR} are for current $Q_T$ measured across the defect bond between $0$-th and $1$-st sites. The generating function \eqref{eq:generating function infinite} corresponds to \eqref{eq:bare G inf} with $\lambda_i=\lambda \,\delta_{i,0}$. Rest of our discussions will be for this particular case, for which, a Gauge transformation $\widehat{n}_{i}\to \widehat{n}_{i}-\lambda\mathbb{I}_{i\ge1}$ (where $\mathbb{I}_{i\ge 1}=1$ for $i\geq1$, otherwise zero) reduces the action in \eqref{gen_fun_inf_expr3 v1} to a convenient form \cite{2009_Derrida_Current2,2022_Mallick_Exact}
\begin{subequations}\label{gen_fun_inf_expr3 action}
\begin{align}
\mathcal{K}&=\lambda \sum_{i\ge 1} (n_i(T)-n_i(0))+\sum_{i}\Bigg\{n_i(0)\,\widehat{n}_i(0)-n_i(T)\,\widehat{n}_i(T)+\int_0^T\dd \tau\,n_i(\tau)\,\frac{\dd \widehat{n}_i(\tau)}{\dd \tau}\Bigg\}\\
\mathcal{H}&=\sum_{i}\,p_i\,\int_0^T\dd \tau\,\omega\big(\widehat{n}_{i+1}(\tau)-\widehat{n}_i(\tau),n_i(\tau),n_{i+1}(\tau)\big)
\end{align}
\end{subequations}

\subsection{Coarse-graining for uniform jump rates ($\alpha=\gamma=1$)\label{sec:coarse graining uniform}}
We begin with the well-studied case of uniform jump rates $p_i=1$ for all $i$. For the long time asymptotic of $G_T$ in \eqref{eq:bare G inf} with $\lambda_i=\lambda \delta_{i,0}$, we invoke an idea of local equilibrium for writing \eqref{gen_fun_inf_expr3} in terms of coarse-grained variables. We shall take a heuristic approach for this coarse-graining.

For the diffusive dynamics of SSEP on infinite lattice, the occupation variables in a subregion of large size $\mathcal{L}\gg 1$ spontaneously reaches an \emph{effective} local equilibrium over a time period $\sim\mathcal{L}^2$, where the statistics of local configurations is described, at the leading order in $\mathcal{L}$, by a Bernoulli distribution with a local average density. This suggests that at any given time, probability of configurations of the \emph{entire} system is described, to the leading order in $\mathcal{L}$, by a product of Bernoulli distributions such that for all sites $i$,
\begin{subnumcases}{\label{occup_prob_measure}n_i(\tau)=}
1 & with prob. $\rho\bigg(\dfrac{i}{\mathcal{L}},\dfrac{\tau}{\mathcal{L}^2}\bigg)$\\
0 & with prob. $1-\rho\bigg(\dfrac{i}{\mathcal{L}},\dfrac{\tau}{\mathcal{L}^2}\bigg)$
\end{subnumcases}
characterized by the smoothly varying density field $\rho(x,t)$ (here, $(x, t) \simeq \left(\frac{i}{\mathcal{L}}, \frac{\tau}{\mathcal{L}^2}\right)$) which is correlated over space and time. This is the local equilibrium hypothesis for our problem.

For the current statistics \eqref{gen_fun_inf_expr3} at large $T$ on the infinite line SSEP we choose $\mathcal{L}\equiv \sqrt{T}$ which is the largest relevant length scale for the problem. For \eqref{gen_fun_inf_expr3}, this local equilibrium assumption also necessitates a smoothly varying conjugate response field
\begin{equation}\label{response_hyd_limit}
\widehat{n}_i(\tau)=\widehat{\rho}\bigg(\frac{i}{\mathcal{L}},\frac{\tau}{\mathcal{L}^2}\bigg)\quad \text{for large $\mathcal{L}\equiv \sqrt{T}$}.
\end{equation}
Errors arising from this local-equilibrium assumption contribute only in the sub-leading order in large $T$ for $G_T$. 

Averaging over this local equilibrium distribution, the generating function \eqref{gen_fun_inf_expr3}, in the large $T$ limit, can be expressed in terms of a path-integral over smooth variables,
\begin{subequations}
\begin{equation}\label{gen_fun_inf_expr4}
G_T\simeq\int_{\boldsymbol{\rho}(0)}^{\boldsymbol{\rho}(1)}\left[\mathcal{D}\rho\right]\left[\mathcal{D}\widehat{\rho}\right]e^{\sqrt{T}\int\dd x\{\rho(x,0)\,\widehat{\rho}(x,0)-\rho(x,1)\,\widehat{\rho}(x,1)\}}\;\mathbb{E}_{\rho(x,t)}[\e^{\mathcal{S}}]
\end{equation}
where we have used \eqref{gen_fun_inf_expr3 action} to write
\begin{equation}\label{eq:micro_action}
\mathcal{S}=\lambda \sum_{i\ge 1} (n_i(T)-n_i(0))+\int_0^T\!\!\!\dd \tau\,\sum_{i}\Bigg\{n_i(\tau)\,\frac{\dd \widehat{n}_i(\tau)}{\dd \tau}+\omega\big(\widehat{n}_{i+1}(\tau)-\widehat{n}_i(\tau),n_i(\tau),n_{i+1}(\tau)\big)\Bigg\}
\end{equation}
with $p_i=1$ for all $i$. Here $\mathbb{E}_\rho$ denotes averaging over microscopic variables $\mathbf{n}$ with the distribution \eqref{occup_prob_measure} for a given  density profile $\rho(x,t)$.
\end{subequations}
In writing \eqref{gen_fun_inf_expr4} we made an additional assumption that, the initial and final configurations, $\mathbf{n}(0)$ and $\mathbf{n}(T)$, are typical with respect to the smooth density $\rho(x,0)$ and $\rho(x,1)$, respectively. Equivalently, $\sum_in_i(0)\,\widehat{n}_i(0)\simeq\sqrt{T}\int\dd x\rho(x,0)\,\widehat{\rho}(x,0)$ and analogous for the final configuration. 

In the local equilibrium assumption \eqref{occup_prob_measure}, temporal correlations of $n_i(\tau)$ are effectively encoded in terms of correlations of smooth fields $\rho(x,t)$. Therefore, for a given history $\rho(x,t)$, the configuration average
\begin{equation}\label{eq:eS ave}
\mathbb{E}_{\rho}\left[\e^{\mathcal{S}}\right]=\mathbb{E}_{\rho}\left[\prod_{\tau}\e^{\dd\tau\, s(\tau)}\right] \simeq \prod_{\tau}\mathbb{E}_{\rho}\left[\e^{\dd\tau\, s(\tau)}\right] \simeq \prod_{\tau}\e^{\dd\tau\, \mathbb{E}_{\rho}\left[s(\tau)\right]} =\e^{\mathbb{E}_{\rho}\left[\mathcal{S}\right]}\quad \textrm{for large $T$,}
\end{equation}
where we denoted $\mathcal{S} =\int \dd \tau s(\tau)$. The first $\simeq$ in \eqref{eq:eS ave} is from independence of configurations over time due to the local equilibrium assumption and the second $\simeq$ is in the vanishing $d\tau$ limit.

For computing $\mathbb{E}_{\rho}\left[\mathcal{S}\right]$ with \eqref{eq:micro_action} we see that the simplest terms
\begin{align}\label{eq:kinetic term average}
\mathbb{E}_{\rho} &\left[\lambda \sum_{i\ge 1} (n_i(T)-n_i(0))+\int_0^T\dd\tau\sum_{i}n_i(\tau)\,\frac{\dd \widehat{n}_i(\tau)}{\dd\tau}\right]\cr
& \qquad \simeq \sqrt{T}\left\{\lambda \int_{0}^{\infty}\dd x\,(\rho(x,1)-\rho(x,0))+\int_0^1\dd t\int_{-\infty}^{\infty}\dd x\,\rho(x,t)\partial_t\widehat{\rho}(x,t)\right\}
\end{align}
scales as $\sqrt{T}$ for large $T$. 

For computing average of the rest of the terms in \eqref{eq:micro_action}, we rewrite the expression of $\omega$ in \eqref{omega_param_defn}: 
\begin{equation}
\omega\big(\widehat{n}_{i+1}-\widehat{n}_i,n_i,n_{i+1}\big)=\big(\,n_i-n_{i+1}\big)\sinh\big(\widehat{n}_{i+1}-\widehat{n}_i\big)+\big(n_i-2n_i n_{i+1}+n_{i+1}\big)2\sinh^2\left(\frac{\widehat{n}_{i+1}-\widehat{n}_i}{2}\right)
\end{equation}
Considering that for the smooth function \eqref{response_hyd_limit} the gradient $\widehat{n}_{i+1}-\widehat{n}_i\sim\frac{1}{\sqrt{T}}$ we get the leading asymptotics
\begin{equation}\label{eq:bulk term average}
\mathbb{E}_{\rho}\left[\int_0^T\dd\tau\sum_{i}\omega\left(\widehat{n}_{i+1}-\widehat{n}_i,n_i,n_{i+1}\right)\right]\simeq \sqrt{T}\int_0^1\dd t\int_{-\infty}^{\infty}\dd x\,\big(\rho(1-\rho)\partial_x\widehat{\rho}-\partial_x\rho\big)\partial_x\widehat{\rho}
\end{equation}

Collecting these results (\ref{eq:kinetic term average}, \ref{eq:bulk term average}) for the local equilibrium averages, we write the long time asymptotic of \eqref{gen_fun_inf_expr4} in terms of the coarse-grained variables
\begin{subequations}\label{eq:mft for uniform inf sep}
\begin{equation}\label{gen_fun_inf_expr6}
G_T\simeq\int_{\rho(x,0)}^{\rho(x,1)}\left[\mathcal{D}\rho\right]\left[\mathcal{D}\widehat{\rho}\right]\e^{-\sqrt{T}\,S\left(\rho,\widehat{\rho}\right)}
\end{equation}
where the action has a Lagrangian structure
\begin{equation}\label{eq:final_action}
S\left(\rho,\widehat{\rho}\right)=-\lambda \int_{0}^{\infty}\dd x\,(\rho(x,1)-\rho(x,0))+\int_0^1\dd t\,\left\{\int_{-\infty}^\infty\dd x\, \left(\widehat{\rho}\,\partial_t\rho\right) -H(\rho,\widehat{\rho})\right\}
\end{equation}
with an effective Hamiltonian
\begin{align}\label{eq:uni h3 parts one}
H(\rho,\widehat{\rho})\equiv H_{\text{uni}}(\rho,\widehat{\rho})&=\int_{-\infty}^{\infty}\dd x\,\Big(\rho(1-\rho)\partial_x\widehat{\rho}-\partial_x\rho\Big)\partial_x\widehat{\rho}.
\end{align}
\end{subequations}
In writing the $\widehat{\rho}\,\partial_t\rho$ term in \eqref{eq:final_action} we have used an integration by parts in time of the expression in \eqref{eq:kinetic term average} to cancel the boundary terms in \eqref{gen_fun_inf_expr4}. For further details about coarse-graining in general stochastic dynamics see \cite{2026_Saha}. The path integral formulation \eqref{eq:mft for uniform inf sep} is the well-known MFT description \cite{2009_Derrida_Current2} for SSEP on infinite line. 

\subsection{Coarse-graining in the presence of a slow bond}
We consider the inhomogeneous case where the jump rate $\gamma$ across the single bond at the origin indicated in Fig.~\ref{fig:inf_ssep_derive} differs from the unit rates ($\alpha=1$) for the rest of the system. The system may be viewed as two homogeneous semi-infinite subsystems with unit jump rate interlinked by a single bond of jump rate $\gamma$.

For writing a coarse-grained description analogous to \eqref{eq:mft for uniform inf sep} we assume that sites on either sides of the defect bond are independently in local equilibrium \eqref{occup_prob_measure} with a coarse-grained density profile $\rho(x,t)$ that is discontinuous across the bond. We denote $\rho_{\pm}(t)\equiv \lim_{x\to\pm 0} \rho(x,t)$, and for the corresponding conjugate fields $\widehat{\rho}_{\pm}(t)\equiv \lim_{x\to \pm0}\widehat{\rho}(x,t)$. 

Extending the coarse-graining steps in Sec.~\ref{sec:coarse graining uniform} by incorporating this discontinuity we get a path integral representation of $G_T$ that is analogous to \eqref{gen_fun_inf_expr6} with an action \eqref{eq:final_action} and a modified effective Hamiltonian
\begin{align}\label{eq:h3 parts one}
H(\rho,\widehat{\rho})\equiv H_{\rm inhom}(\rho,\widehat{\rho})&=\int_{-\infty}^0\dd x\,\Big[\rho(1-\rho)\partial_x\widehat{\rho}-\partial_x\rho\Big]\partial_x\widehat{\rho}+\Gamma\,\omega\big(\widehat{\rho}_{+}(t)-\widehat{\rho}_{-}(t),\rho_{-}(t),\rho_{+}(t)\big)\nonumber\\
&\quad+\int_0^\infty\dd x\,\Big[\rho(1-\rho)\partial_x\widehat{\rho}-\partial_x\rho\Big]\partial_x\widehat{\rho},
\end{align}
where we denote $\Gamma=\gamma\sqrt{T}$. 

The term involving $\Gamma$ in \eqref{eq:h3 parts one} shows that maintaining the discontinuity in $\rho(x,t)$ and $\widehat{\rho}(x,t)$ across the origin is costly for large $T$ unless $\gamma$ is as small as $\sim \frac{1}{\sqrt{T}}$. In the following we shall consider this marginal case.

\subsection{Current statistics across the slow bond with marginal jump rate}
The result presented in \eqref{eq:mu_slow_inf} is for current $Q_T$ across the slow bond with jump rate $\gamma=\frac{\Gamma}{\sqrt{T}}$. The generating function
\begin{equation}\label{eq:G_T}
\left\langle\e^{\lambda Q_T}\right\rangle = \sum_{\boldsymbol{n}(0)}\sum_{\boldsymbol{n}(T)} P(\boldsymbol{n}(0))\,G_T\!\left(\{\lambda \delta_{i,0}\}\vert  \boldsymbol{n}(0),\boldsymbol{n}(T)\right)
\end{equation}
where $G_T$ is defined in \eqref{eq:bare G inf} and $P(\boldsymbol{n}(0))$ is the distribution in the initial state.

Initially, the system is prepared in a domain-wall state where $i\le 0$ sites are independently occupied with Bernoulli distribution of uniform average density $\rho_a$, while the $i>0$ sites with average density $\rho_b$, accordingly. The corresponding probability distribution of the initial density profile $\rho(x,0)$ has a large deviation asymptotic $P(\rho(x,0))\sim\mathrm{e}^{-\sqrt{T}\,\mathcal{F}((\rho(x,0))}$ with \cite{2007_Derrida_Non,2025_Derrida_Les_Houches},
\begin{equation}\label{eq:Peq}
\mathcal{F}(\rho(x))=\int_{-\infty}^0\dd x \, f(\rho(x),\rho_a)+\int_{0}^\infty\dd x \, f(\rho(x),\rho_b)
\end{equation}
where
\begin{equation}
f(\rho,\bar{\rho})= \rho\log\frac{\rho}{\bar{\rho}} + (1-\rho)\log\frac{1-\rho}{1-\bar{\rho}}.
\end{equation}

Incorporating \eqref{gen_fun_inf_expr6} and \eqref{eq:Peq} we find that the generating function \eqref{eq:G_T} has the large $T$ asymptotic
\begin{equation}\label{eq:full gen fnc MFT}
\langle\e^{\lambda Q_T}\rangle \simeq\int\left[\mathcal{D}\rho\right]\left[\mathcal{D}\widehat{\rho}\right]\e^{-\sqrt{T}\,\left(S\left(\rho,\widehat{\rho}\right)+\mathcal{F}(\rho(x,0))\right)}
\end{equation}
with $S$ in \eqref{eq:final_action} and the effective Hamiltonian \eqref{eq:h3 parts one}.

\subsubsection*{Scaled cumulant generating function}

In the large $T$ limit, the leading asymptotic of the generating function \eqref{eq:full gen fnc MFT} is dominated by the saddle point which leads to the large deviation asymptotic similar to \eqref{eq:generating function infinite} with the scgf (see also notation in \eqref{eq:mu_slow_inf})
\begin{equation}\label{eq:mu optimal first}
\mu_{\text{inf}}^{\text{slow}}(\lambda,\rho_a,\rho_b,\Gamma,1)=\max_{\rho}\min_{\widehat{\rho}}\big\{-S(\rho,\widehat{\rho})-\mathcal{F}(\rho(x,0))\big\},
\end{equation}
where the action
\begin{equation}\label{eq:final_action_inhom}
S(\rho,\widehat{\rho})=-\lambda \int_{0}^{\infty}\dd x\,(\rho(x,1)-\rho(x,0))+\int_0^1\dd t\,\bigg\{\int_{-\infty}^\infty\dd x\,\big(\widehat{\rho}\,\partial_t\rho\big)-H_{\rm inhom}(\rho,\widehat{\rho})\bigg\}
\end{equation}
with $H_{\rm inhom}$ in \eqref{eq:h3 parts one} and $\mathcal{F}$ in \eqref{eq:Peq}. The optimisation is with boundary condition 
\begin{subnumcases}
{(\rho(x,t),\widehat{\rho}(x,t))=}
(\rho_a,0) & for $x\to-\infty$\\
(\rho_b,\lambda) & for $x\to \infty$
\end{subnumcases}
at all times $t\ge 0$.

The subtle reason for taking a minimum, rather than a maximum, over $\widehat{\rho}$ in \eqref{eq:mu optimal first} is that this field is, by construction, purely imaginary (see \eqref{eq:nhat intro}). Consequently, within the steepest-descent method \cite{DenneryKrzywicki2012}, the path-integral is dominated by a saddle point of $-S(\rho,\widehat{\rho})-\mathcal{F}(\rho(x,0))$, rather than by a local maximum in the $(\rho,\widehat{\rho})$-space. This is one the rare instances in MFT where the complex nature of the conjugate field becomes relevant.

For explicitly solving the variational problem \eqref{eq:mu optimal first}, we consider the system composed of three parts: left semi-infinite domain $(x<0)$, right semi-infinite domain $(x>0)$, and the central slow bond connecting the two semi-infinite domains. Following this compartmentalisation we write
\begin{equation}
-S-\mathcal{F}=\mathbb{L}+\Gamma \int_0^1 \dd t \,\omega\big(\widehat{\rho}_{+}(t)-\widehat{\rho}_{-}(t),\rho_{-}(t),\rho_{+}(t)\big) +\mathbb{R}
\end{equation}
with
\begin{equation}
\mathbb{L}=-\int_{-\infty}^0\dd x \int_0^1\dd t\,\bigg\{\widehat{\rho}\,\partial_t\rho-\Big(\rho(1-\rho)\partial_x\widehat{\rho}-\partial_x\rho\Big)\partial_x\widehat{\rho}\bigg\}-\int_{-\infty}^0\dd x \, f(\rho(x,0),\rho_a)
\end{equation}
and
\begin{equation}
\mathbb{R}=\lambda\int_{0}^{\infty}\dd x\,(\rho(x,1)-\rho(x,0))-\int_{0}^\infty\dd x\,\int_0^1\dd t\,\bigg\{\widehat{\rho}\,\partial_t\rho-\Big(\rho(1-\rho)\partial_x\widehat{\rho}-\partial_x\rho\Big)\partial_x\widehat{\rho}\bigg\}-\int_{0}^{\infty}\dd x\, f(\rho(x,0),\rho_b)\label{eq:Rbb explicit}
\end{equation}

With this compartmentalisation we carry out the optimization \eqref{eq:mu optimal first} in two steps, first in the two semi-infinite sub-systems while keeping the boundary fields $(\rho_{\pm}(t),\widehat{\rho}_{\pm}(t))$ fixed, which are subsequently optimised incorporating the contributions from the central slow bond.
We further make an additivity ansatz \cite{2004_Bodineau_Current}, that the optimal boundary fields $(\rho_{\pm}(t),\widehat{\rho}_{\pm}(t))$ are constant for most part of the time period $t\in [0,1]$. 

This reduces the optimisation \eqref{eq:mu optimal first} to
\begin{equation}\label{eq:mu inf slow one LR}
\mu_{\text{inf}}^{\text{slow}}(\lambda,\rho_a,\rho_b,\Gamma,1)\simeq\max_{\rho_\pm}\min_{\widehat{\rho}_\pm}\bigg\{\mathbb{L}_{\rm opt}(\widehat{\rho}_-,\rho_a,\rho_-)+\Gamma\,\omega\big(\widehat{\rho}_+-\widehat{\rho}_-,\rho_-,\rho_+\big)+\mathbb{R}_{\rm opt}(\lambda,\widehat{\rho}_+,\rho_+,\rho_b)\bigg\}
\end{equation}
where $(\rho_{\pm},\widehat{\rho}_{\pm})$ are constant and we denote 
\begin{equation}\label{eq:Lopt}
\mathbb{L}_{\rm opt}(\widehat{\rho}_-,\rho_a,\rho_-)=\max_{\rho}\min_{\widehat{\rho}}\mathbb{L}
\end{equation}
in the negative $x$ domain with the time-independent boundary condition
\begin{subnumcases}
{(\rho,\widehat{\rho})=}
(\rho_a,0) & for $x\to-\infty$\\
(\rho_-,\widehat{\rho}_-) & for $x\to0^-$
\end{subnumcases}
while
\begin{equation}\label{eq:Ropt}
\mathbb{R}_{\rm opt}(\lambda,\widehat{\rho}_+,\rho_+,\rho_b)=\max_{\rho}\min_{\widehat{\rho}}\mathbb{R}
\end{equation}
in the positive $x$ domain with the time-independent boundary condition
\begin{subnumcases}
{(\rho,\widehat{\rho})=}
(\rho_+,\widehat{\rho}_+) & for $x\to0^+$\\
(\rho_b,\lambda) & for $x\to\infty$.
\end{subnumcases}

Using the explicit expression for $\mathbb{R}$ in \eqref{eq:Rbb explicit} and a transformation $\widehat{\rho}(x,t)\to \widehat{\rho}(x,t)-\widehat{\rho}_+$, it is simple (see \cite{2026_Sharma_Large}) to see that $\mathbb{R}_{\rm opt}(\lambda,\widehat{\rho}_+,\rho_+,\rho_b)$ in \eqref{eq:Ropt} is the scgf $\mu_{\rm si}(\lambda-\widehat{\rho}_+,\rho_+,\rho_b)$ of current with fugacity $\lambda-\widehat{\rho}_+$ exchanged between reservoir at density $\rho_+$ and a semi-infinite ($\mathbb{Z^+}$) SSEP initially at an average uniform density $\rho_b$. Solving the corresponding variational problem for the semi-infinite SSEP is challenging in itself and has been explicitly solved in \cite{2026_Sharma_Large}
providing
\begin{equation}\label{eq:mu si 52}
\mu_{\text{si}}(\lambda,\rho_a,\rho_b)= R_{\text{si}}\big(\omega(\lambda,\rho_a,\rho_b) \big)
\end{equation}
with $R_{\rm si}$ in \eqref{PW_Rsi}.

A similar analysis shows that $\mathbb{L}_{\rm opt}(\widehat{\rho}_-,\rho_a,\rho_-)$ in \eqref{eq:Lopt} is the scgf $\mu_{\rm si}(-\widehat{\rho}_-,\rho_-,\rho_a)$ of current with fugacity $(-\widehat{\rho}_{-})$ exchanged between a reservoir at density $\rho_-$ and a semi-infinite SSEP with initial density $\rho_a$. 

This relation to the semi-infinite SSEP solves the first step of the variational problem \eqref{eq:mu optimal first} leading to
\begin{equation}
\mu_{\text{inf}}^{\text{slow}}(\lambda,\rho_a,\rho_b,\Gamma,1)\simeq\max_{\rho_\pm}\min_{\widehat{\rho}_\pm}\Big\{R_{\text{si}}\big(\omega(\widehat{\rho}_-,\rho_a,\rho_-)\big)+\Gamma\,\omega\big(\widehat{\rho}_+-\widehat{\rho}_-,\rho_-,\rho_+\big)+R_{\text{si}}\big(\omega(\lambda-\widehat{\rho}_+,\rho_+,\rho_b)\big)\Big\}\label{eq: 4d_variational}
\end{equation}
The remaining optimisation can be done following an identity given in \cite{2021_Derrida_Large},
\begin{align}\label{eq: variational_id}
\max_{\rho}\min_{\lambda} \left\{A(\omega(\lambda, \rho_1,\rho))+B(\omega(\lambda_1-\lambda, \rho, \rho_2))\right\}=\min_{z}\left\{A(\sinh^2 z) + B(\sinh^2(z\pm u))\right\}
\end{align}
where $\omega(\lambda_1, \rho_1, \rho_2)=\sinh^2u$. Applying this identity twice reduces the 4-dimensional variational problem~\eqref{eq: 4d_variational} to the following simpler expression 
\begin{equation}
\mu_{\text{inf}}^{\text{slow}}(\lambda,\rho_a,\rho_b,\Gamma,1)=R_{\text{inf}}^{\text{slow}}(\omega(\lambda,\rho_a,\rho_b),\Gamma,1)
\end{equation}
with
\begin{equation}\label{eq:variational infinite ssep single}
R_{\text{inf}}^{\text{slow}}(\omega,\Gamma,1)=\min_{z_a, z_b}\Bigg\{R_{\text{si}}\big(\sinh^2{z_a}\big)+\Gamma \sinh^2{(z_a+z_b-\text{arcsinh}\,\sqrt{\omega})}+R_{\text{si}}\big(\sinh^2{z_b}\big)\Bigg\}
\end{equation}
which is consistent with the more general expression \eqref{eq:variational infinite ssep} we reported.

\textit{Remark:} Analysis for the case of multiple slow bonds involves an MFT action with additional terms in the Hamiltonian $H_0$ which are easy to construct following our method. For example, for two slow bonds between the $(-1)$th and the $(+1)$th sites, 
\begin{equation}
H_0=\Gamma\,\omega\big(\widehat{\rho}_{0}-\widehat{\rho}_{-},\rho_{-},\rho_{0}\big)+\Gamma\,\omega\big(\widehat{\rho}_{+}-\widehat{\rho}_{0},\rho_{0},\rho_{+}\big)
\end{equation}
where $\rho_0(t)$ is the density at the $0$-th site. The analysis can be done following a similar approach leading to the result \eqref{eq:variational infinite ssep}.

\section{The semi-infinite SSEP weakly coupled to a reservoir \label{sec:si full}}
The current statistics for the semi-infinite SSEP in Fig.~\ref{fig:semi_inf_with_slow_boundary_coupling} can be simply understood from the $\alpha\to \infty$ of the corresponding problem on infinite SSEP in Fig.~\ref{fig:inf_ssep_derive}. This corresponds to the scenario where the dynamics is significantly faster in the left-half of the system, which acts as a particle reservoir for the right half of the system.

In this limit, fluctuations on the negative $x$ domain are suppressed with $\rho(x,t)=\rho_a$ and $\widehat{\rho}(x,t)=0$ for $x<0$. This reduces the generating function of current \eqref{eq:full gen fnc MFT} to 
\begin{equation}\label{eq:full gen fnc semi inf}
\langle\e^{\lambda Q_T}\rangle \simeq\int\left[\mathcal{D}\rho\right]\left[\mathcal{D}\widehat{\rho}\right]\e^{-\sqrt{T}\,\left(S_{\rm si}\left(\rho,\widehat{\rho}\right)+\int_{0}^\infty\dd x \, f(\rho(x,0),\rho_b)\right)}
\end{equation}
with the action
\begin{align}
S_{\rm si}\left(\rho,\widehat{\rho}\right)=&-\lambda \int_{0}^{\infty}\dd x\,(\rho(x,1)-\rho(x,0))-\Gamma \int_0^1 \dd t \,\omega\big(\widehat{\rho}_{+}(t),\rho_a,\rho_{+}(t)\big)\nonumber\\
&+\int_{0}^\infty\dd x\,\int_0^1\dd t\,\bigg\{\widehat{\rho}\,\partial_t\rho-\Big(\rho(1-\rho)\partial_x\widehat{\rho}-\partial_x\rho\Big)\partial_x\widehat{\rho}\bigg\}
\end{align}

The large $T$ asymptotic $\big<\e^{\lambda Q_T}\big>\simeq\e^{\sqrt{T}\mu_{\text{si}}^{\text{slow}}(\lambda,\rho_a,\rho_b)}$ comes from a saddle point solution of \eqref{eq:full gen fnc semi inf} yielding
\begin{equation}
\mu_{\text{si}}^{\text{slow}}(\lambda,\rho_a,\rho_b)\simeq\max_{\rho}\min_{\widehat{\rho}}\left\{-S_{\rm si}\left(\rho,\widehat{\rho}\right)-\int_{0}^\infty\dd x \, f(\rho(x,0),\rho_b)\right\}
\end{equation}
with a spatial boundary condition $(\rho,\widehat{\rho})=(\rho_b,\lambda)$ for $x\to \infty$.

This optimisation problem is solved in parts following steps similar to the infinite line case in \eqref{eq:mu optimal first}, which leads to
\begin{equation}
\mu_{\text{si}}^{\text{slow}}(\lambda,\rho_a,\rho_b)\simeq\max_{\rho_0}\min_{\widehat{\rho}_0}\left\{\Gamma\,\omega\big(\widehat{\rho}_0,\rho_a,\rho_0\big)+\mu_{\text{si}}(\lambda-\widehat{\rho}_0,\rho_0,\rho_b)\right\}
\end{equation}
with the semi-infinite scgf $\mu_{\rm si}$ in the fast coupling limit in \eqref{eq:mu si 52}.

The expression further reduces to $\mu_{\text{si}}^{\rm slow}(\lambda,\rho_a,\rho_b)=R_{\text{si}}^{\text{slow}}(\omega(\lambda,\rho_a,\rho_b),\Gamma)$ using the identity \eqref{eq: variational_id} leading to the reported result \eqref{eq:var si}.

\section{The finite SSEP with two slowly coupled reservoirs \label{sec:fin full}}

In this Section, we outline the main steps for the derivation of the result \eqref{eq:var sol fin} about the current statistics in a finite SSEP slowly coupled with boundary reservoirs. We follow an analysis similar to the infinite-line case in Sec.~\ref{sec:inf SSEP}. A detailed discussion can be found in our earlier publication \cite{2024_Saha_Large}. 

The problem is defined on a finite one-dimensional lattice with $L$ sites, sketched in Fig.~\ref{fig:fin_with_slow_boundary_coupling}. Within the bulk, the hopping rates are taken as unity. The lattice is connected to two boundary reservoirs, through slow bonds, and modelled by particle exchanges with rates indicated in Fig.~\ref{fig:fin_with_slow_boundary_coupling}. 

Across the bonds in the bulk, probability of the jump variables $Y_i$ for $1\le i\le L-1$ is given in \eqref{lft_semi_inf}. Across the bonds connecting the lattice to the left and right reservoirs, their probability is the following.
\begin{subnumcases}
{\label{hop_prob_lft_bdry_fin}Y_0(\tau)=}
1 & with prob. $\gamma\rho_a\,\big(1-n_1(\tau)\big)\,\dd \tau$\\
-1 & with prob. $\gamma n_1(\tau)\,\big(1-\rho_a\big)\,\dd \tau$\\
0 & with prob. $1-\gamma\big(\rho_a\big(1-n_1(\tau)\big)+n_1(\tau)\,\big(1-\rho_a\big)\big)\,\dd \tau$
\end{subnumcases}
and
\begin{subnumcases}
{\label{hop_prob_rgt_bdry_fin}Y_L(\tau)=}
1 & with prob. $\gamma n_L(\tau)\,\big(1-\rho_b\big)\,\dd \tau$\\
-1 & with prob. $\gamma\rho_b\,\big(1-n_L(\tau)\big)\,\dd \tau$\\
0 & with prob. $1-\gamma\big(\rho_b\big(1-n_L(\tau)\big)+n_L(\tau)\,\big(1-\rho_b\big)\big)\,\dd \tau$
\end{subnumcases}
respectively.

Following steps described in the infinite line problem in Sec.~\ref{sec:inf SSEP}, we write the generating function for $Q_i(T)$,
\begin{align}
G_T&\left(\boldsymbol{\lambda}\vert \boldsymbol{n}(0),\boldsymbol{n}(T)\right):=\Big<\e^{\sum_i\lambda_iQ_i(T)}\,\big\vert\,\boldsymbol{n}(T),\boldsymbol{n}(0)\Big>\cr & =\int_{\boldsymbol{n}(0)}^{\boldsymbol{n}(T)}\left[\mathcal{D}n\right]\Bigg<\e^{\sum_{k=0}^{M-1}\sum_{i=0}^L\lambda_iY_i(k\,\dd \tau)}\prod_{k=0}^{M-1}\,\prod_{i=1}^L\Big(\delta_{n_i(k\,\dd \tau+\dd \tau)-n_i(k\,\dd \tau),Y_{i-1}(k\,\dd \tau)-Y_i(k\,\dd \tau)}\Big)\Bigg>_{\boldsymbol{Y}}\label{gen_fun_fin_expr1}
\end{align}
which, using the integral representation of the Kronecker-delta function, gives
\begin{align}
G_T=&\int_{\boldsymbol{n}(0)}^{\boldsymbol{n}(T)}\left[\mathcal{D}n\right]\left[\mathcal{D}\widehat{n}\right]\exp{\bigg[-\sum_{k=0}^{M-1}\,\sum_{i=1}^L\widehat{n}_i(k\,\dd \tau)\,\big(n_i(k\,\dd \tau+\dd \tau)-n_i(k\,\dd \tau)\big)\bigg]}\nonumber\\
&\quad\Bigg<\exp{\bigg(\sum_{k=0}^{M-1}\,\sum_{i=1}^{L-1}\big(\widehat{n}_{i+1}(k\,\dd \tau)-\widehat{n}_i(k\,\dd \tau)+\lambda_i\big)\,Y_i(k\,\dd \tau)\bigg]}\exp{\bigg[\sum_{k=0}^{M-1}\widehat{n}_1(k\,\dd \tau)\,Y_0(k\,\dd \tau)
\bigg)}\nonumber\\
&\qquad\qquad\qquad\exp{\bigg(-\sum_{k=0}^{M-1}\widehat{n}_L(k\,\dd \tau)\,Y_L(k\,\dd \tau)\bigg)}\Bigg>_{\boldsymbol{Y}}\label{gen_fun_fin_expr2}
\end{align}
In rewriting the above expression for the generating function, we have made use of the identity, $\sum_{i=1}^L\widehat{n}_i\,\big(Y_{i-1}-Y_i\big)=\sum_{i=1}^{L-1}\big(\widehat{n}_{i+1}-\widehat{n}_i\big)\,Y_i+\widehat{n}_1\,Y_0
-\widehat{n}_L\,Y_L$, where the first term involving $Y_i$ with $1\le i\le L-1$ comes from hopping events across the bonds in the bulk, whereas $Y_0$ ($Y_L$) involves the hopping across the slow bond connecting the lattice to the left (right) reservoir. Performing the averaging over the hopping events in the bulk part using \eqref{lft_semi_inf} and across the bond connected with the left and right reservoirs using (\ref{hop_prob_lft_bdry_fin}-\ref{hop_prob_rgt_bdry_fin}), we obtain from \eqref{gen_fun_fin_expr2}
\begin{subequations}\label{gen_fun_fin_expr3}
\begin{equation}
G_T=\int_{\boldsymbol{n}(0)}^{\boldsymbol{n}(T)}\left[\mathcal{D}n\right]\left[\mathcal{D}\widehat{n}\right]\e^{\mathcal{K}+\mathcal{H}}
\end{equation}
with the exponential terms given by
\begin{align}
\mathcal{K}&=\sum_{i=1}^L\Bigg\{n_i(0)\,\widehat{n}_i(0)-n_i(T)\,\widehat{n}_i(T)+\int_0^{T}\dd \tau\,n_i(\tau)\,\frac{\dd \widehat{n}_i(\tau)}{\dd \tau}\Bigg\}\\
\mathcal{H}&=\gamma\int_0^{T}\dd \tau\,\omega\big(\widehat{n}_1(\tau)+\lambda_0,\rho_a,n_1(\tau)\big)+\gamma\int_0^{T}\dd \tau\,\omega\big(\widehat{n}_L(\tau)-\lambda_L,\rho_b,n_L(\tau)\big)\cr&\qquad+\sum_{i=1}^{L-1}\int_0^{T}\dd \tau\,\omega\big(\widehat{n}_{i+1}(\tau)-\widehat{n}_i(\tau)+\lambda_i,n_i(\tau),n_{i+1}(\tau)\big)
\end{align}
\end{subequations}
with the $\omega$-function defined in \eqref{omega_param_defn}.

For our analysis we consider $\lambda_i=\lambda \delta_{i,0}$ which corresponds to the current measured at the bond between the left reservoir and the system. For the finite SSEP, the current statistics at large time $T\gg L^2$ is independent of the location where current is measured \cite{Derrida2019LargeDeviationsII,2024_Saha_Large}. This can be seen following a Gauge transformation $\widehat{n}_i\to \widehat{n}_i-\sum_{j\ge i}^{L}\lambda_j$ in \eqref{gen_fun_fin_expr3}.

For a large system size $L$ compared to the coarse graining length $\mathcal{L}\gg 1$, we evaluate the leading asymptotic of generating function in \eqref{gen_fun_fin_expr3} by expressing the path integral in terms of coarse-grained variables $n_i(\tau)\simeq \rho \left(\frac{i}{\mathcal{L}}, \frac{\tau}{\mathcal{L}^2} \right)$ and corresponding conjugate field $\widehat{n}_i(\tau)\simeq \widehat{\rho} \left(\frac{i}{\mathcal{L}}, \frac{\tau}{\mathcal{L}^2} \right)$. The coarse-graining is done following the heuristic procedure discussed in the infinite line problem in Sec.~\ref{sec:coarse graining uniform} that leads to the large $\mathcal{L}$ asymptotics
\begin{align}
G_T\simeq \int\left[\mathcal{D}\rho\right]\left[\mathcal{D}\widehat{\rho}\right]\exp\Bigg\{-\mathcal{L}\int_0^{\frac{T}{{\mathcal{L}}^2}}\mathrm{d}t\,\Bigg[&\int_0^{\ell}\mathrm{d}x\,\big(\widehat{\rho}\partial_t\rho\big)-H(\rho,\widehat{\rho})\Bigg]\Bigg\}\label{eq:path integral current}
\end{align}
with an effective Hamiltonian
\begin{equation}
H(\rho,\widehat{\rho})=\frac{\Gamma}{\ell}\omega\big(\widehat{\rho}_{\rm lt}(t)+\lambda,\rho_a,\rho_{\rm lt}(t)\big)+\frac{\Gamma}{\ell} \omega\big(\widehat{\rho}_{\rm rt}(t),\rho_b,\rho_{\rm rt}(t)\big)+\int_0^{\ell}\mathrm{d}x\big[\rho(1-\rho)\partial_x\widehat{\rho}-\partial_x\rho\big]\partial_x\widehat{\rho}
\label{bulk_hamiltonian}
\end{equation}
where we define $\gamma=\frac{\Gamma}{L}$ and $\ell=\frac{L}{\mathcal{L}}$. The subscript `${\rm lt}$' (`${\rm rt}$') denotes the fields at the left (right) boundary.

For large $\mathcal{L}$ the path integral in \eqref{eq:path integral current} is dominated by the saddle point. For large $T$, the optimal path associated to a saddle point stays at a quasi-stationary value for most period of evolution and the leading asymptotic can be computed by a time-independent solution. The resulting asymptotic $G_T\asymp\e^{\frac{T \ell}{L}\chi}$ with
\begin{equation}\label{eq:sep fin opt mu first}
\chi=\max_{\rho}\min_{\widehat{\rho}}\left\{ \frac{\Gamma}{\ell}\omega\big(\widehat{\rho}_{\rm lt}+\lambda,\rho_a,\rho_{\rm lt}\big)+\frac{\Gamma}{\ell}\omega\big(\widehat{\rho}_{\rm rt},\rho_b,\rho_{\rm rt}\big)+\int_0^{\ell}\mathrm{d}x\big[\rho(1-\rho)\partial_x\widehat{\rho}-\partial_x\rho\big]\partial_x\widehat{\rho}\right\}
\end{equation}

The optimisation can be done in parts, first in the bulk while keeping the values of the fields at the boundary fixed. It is simple to see \cite{2024_Saha_Large} that the corresponding optimal value from the bulk in \eqref{eq:sep fin opt mu first} correspond to the scgf of current in the fast coupling limit $(\Gamma\to \infty)$ and has been determined earlier \cite{2004_Bodineau_Current,2004_Derrida_Current}. Using the corresponding explicit result we write
\begin{equation}\label{eq:sep fin opt mu first explicit}
\chi=\max_{\rho_{\rm lt},\rho_{\rm rt}}\min_{\widehat{\rho}_{\rm lt},\widehat{\rho}_{\rm rt}}\left\{ \frac{\Gamma}{\ell}\omega\big(\widehat{\rho}_{\rm lt}+\lambda,\rho_a,\rho_{\rm lt}\big)+\frac{\Gamma}{\ell}\omega\big(\widehat{\rho}_{\rm rt},\rho_b,\rho_{\rm rt}\big)+\frac{1}{\ell}\left(\rm{arcsinh}\sqrt{\omega(\widehat{\rho}_{\rm rt}-\widehat{\rho}_{\rm lt},\rho_{\rm lt},\rho_{\rm rt})}\right)^2\right\}
\end{equation}

This remaining optimisation can be completed using the identity \eqref{eq: variational_id}
leading to
\begin{equation}
\chi =\frac{1}{\ell}\min_{z_a,z_b}\Big\{\Gamma\sinh^2{z_a}+\Gamma\sinh^2{z_b}+\left(z_a+z_b-\text{arcsinh}\,\sqrt{\omega(\lambda,\rho_a,\rho_b)}\right)^2\Big\}
\end{equation}
This gives the asymptotic of the generating function $G_T\asymp e^{\frac{T\ell}{L}\chi}= e^{T R_{L}^{\text{slow}}}$ consistent with the reported result \eqref{eq:var sol fin}.

\section{Fluctuating hydrodynamics \label{sec:fhd full}}
Our analysis for the current statistics is based on a hydrodynamics description for SSEP which is presented in terms of path integral. In the Martin-Siggia-Rose-Janssen-De Dominicis (MSRJD) formalism \cite{1973_Martin_Statistical,1976_Janssen_On,1976_Dominicis_Techniques,1978_Dominicis_Field} these path integrals are associated to a stochastic differential equation for the density fields which is the corresponding fluctuating hydrodynamics. For a SSEP on a homogenous lattice with unit jump rates, the coarse-grained density field $\rho(x,t)\simeq n_i(\tau)$ at hydrodynamic coordinates $(x,t)\equiv \left(\frac{i}{\mathcal{L}},\frac{\tau}{\mathcal{L}^2}\right)$ with large coarse-graining length $\mathcal{L}$ follows 
\begin{equation}\label{eq:fhd homo}
\partial_t\rho(x,t)=\partial_x^2\rho(x,t)+\frac{1}{\sqrt{\mathcal{L}}}\partial_x\left(\sqrt{2\rho(x,t)(1-\rho(x,t))}\; \eta(x,t)\right)
\end{equation}
where $\eta(x,t)$ is a zero-mean Gaussian noise with a covariance that is delta-correlated in space and time and of unit strength. 

Our analysis in this work showed that even in presence of microscopic defects the large scale properties can be captured by a suitable extension of the hydrodynamics description \eqref{eq:fhd homo}. We summarise below this fluctuating hydrodynamics in presence of slow-bonds in the three geometries considered. In all three cases, effect of a localized slow bond is captured by a boundary condition across the slow bond.

\paragraph{SSEP on infinite lattice.} Jump rates indicated in Fig.~\ref{fig:inf_with_l_slow_bonds} are one across every bonds except for a \emph{single} defect bond at origin where the jump rate $\gamma=\frac{\Gamma}{\mathcal{L}}$. Corresponding fluctuating hydrodynamics \eqref{eq:fhd homo} is defined separately for the domains $x<0$ and $x>0$, with a boundary condition
\begin{equation}
\partial_x\rho(0^{\mp},t)=\xi_{\mp}(t)\qquad\textrm{at $x=0^{\mp}$}
\end{equation}
where statistics of the noises $\xi_\mp$ are defined in terms of their generating functions
\begin{equation}
\Big<\e^{\int\dd t\,h(t)\xi_{\mp}(t)}\Big>=\e^{\int\dd t\,\left\{\Gamma\omega\left(h(t),\rho(0^-,t),\rho(0^+,t)\right)-h(t)^2\rho(0^{\mp},t)\left(1-\rho(0^{\mp},t)\right)\right\}}
\end{equation}

\paragraph{SSEP on semi-infinite lattice.} Rates are indicated in Fig.~\ref{fig:semi_inf_with_slow_boundary_coupling} with a slow bond at origin coupling the system to a reservoir. For the jump parameter $\gamma=\frac{\Gamma}{\mathcal{L}}$, corresponding fluctuating hydrodynamics \eqref{eq:fhd homo} is defined on the positive domain $x>0$ with a boundary condition
\begin{equation}
\partial_x\rho(0,t)=\xi_{\rm lt}(t)
\end{equation}
with noise characteristic
\begin{equation}\label{slow_si_bond_noise}
\Big<\e^{\int\dd t\,h(t)\,\xi_{\rm lt}(t)}\Big>=\e^{\int\dd t\,\left\{\Gamma\omega\left(h(t),\rho_a,\rho(0,t)\right)-h(t)^2\rho(0,t)\left(1-\rho(0,t)\right)\right\}}
\end{equation}

\paragraph{SSEP on finite lattice.} Rates are indicated in Fig.~\ref{fig:fin_with_slow_boundary_coupling} with the two slow bonds at the boundary coupling the system to two reservoirs. For the jump parameters $\gamma_{a(b)}=\frac{\Gamma}{\mathcal{L}}$ the corresponding fluctuating hydrodynamics \eqref{eq:fhd homo} is defined on the finite domain $x\in (0,\ell)$ with $\ell=\frac{L}{\mathcal{L}}$ and a boundary condition
\begin{align}
\partial_x \rho(x,t)&=\xi_{\rm lt}(t)\qquad \textrm{at $x=0$,}\\
\partial_x \rho\left(x,t\right)&=\xi_{\rm rt}(t) \qquad \textrm{at $x=\ell$,}
\end{align}
with $\xi_{\rm lt}$ in \eqref{slow_si_bond_noise} and 
\begin{equation}
\Big<\e^{\int\dd t\,h(t)\,\xi_{\rm rt}(t)}\Big>=\e^{\int\dd t\,\left\{\Gamma\omega\left(h(t),\rho(\ell,t),\rho_b\right)-h(t)^2\rho(\ell,t)\left(1-\rho(\ell,t)\right)\right\}}
\end{equation}

\section{Details of numerical simulation \label{sec:simulation}}
The rare-event simulation results for the scgf of the current in Figs.~\ref{fig:inf_slow_cgf}, \ref{fig:semi_inf_slow_cgf}, and \ref{fig: finite_slow_cgf} are based on the continuous-time cloning algorithm \cite{2006_Giardinà_Direct, 2007_Lecomte_Numerical, 2019_Perez_Sampling} for the SSEP. The algorithm provides a direct approach for computing the scgf of time-integrated observables, such as $Q_T$, by numerically evaluating the largest eigenvalue of associated tilted operator. This is done by constructing a Markov process from the non-Markovian tilted operator by introducing cloning and culling of a population of samples. We refer to earlier references \cite{2006_Giardinà_Direct, 2007_Lecomte_Numerical, 2019_Perez_Sampling} for details of the algorithm. Below, we provide relevant details of our numerical simulation in each cases that we reported.

\subsection{SSEP on an Infinite Lattice with Slow Bonds}
The result shown in Fig.~\ref{fig:inf_slow_cgf} is for the SSEP on an infinite lattice with localized slow bonds around the origin. In practice it is not feasible to simulate an infinite lattice. However, the statistics of current $Q_T$ for finite time $T$ can be effectively captured by a finite lattice of appropriate large length \(2L+1\) with reflecting boundary conditions at \(-L\) and \(+L\). Initially, the lattice is populated following a Bernoulli distribution with domain wall average density: \(\rho_a\) and \(\rho_b\) to the left and right of the origin, respectively. With time $T$, the average density profile evolves with the interface width spreading diffusively as $\sqrt{T}$. The statistics near the origin remains largely unaffected by the boundary for time $T\ll L^2$. Therefore, the statistics of empirical current $Q_T$ measured across the origin on a large finite lattice $L\gg \sqrt{T}$ to the leading order in $T$ captures the corresponding statistics on an infinite lattice. 

In practice, we find that for uniform density \((\rho_a, \rho_b) = \left(\frac{1}{2}, \frac{1}{2}\right)\) a system of size $L=50$ captures the current statistics for $T=500$ on the infinite-lattice as shown in Fig.~\ref{fig:inf_slow_cgf}. The result is averaged over a clone population $N_c=10^4$. Noticeable deviations in Fig.~\ref{fig:inf_slow_cgf} at large fluctuations are due to the finite size effects from the system length, time, and the clone population. We consider the case where hopping rate across the slow bonds is $\gamma=\frac{\Gamma}{\sqrt{T}}$, while it is \(1\) elsewhere, as illustrated in Fig.~\ref{fig:inf_with_l_slow_bonds}. We presented results for different number of slow bonds and for two sets of $\Gamma=(1,2)$.

\subsection{SSEP on semi-infinite lattice}
For simulation of semi-infinite lattice we follow a similar approach by considering finite lattice of length $L$, coupled to a reservoir on the left most site while imposing a reflecting boundary condition on the right-most site. Jump rates across bonds are illustrated in Fig.~\ref{fig:semi_inf_with_slow_boundary_coupling}. The empirical current $Q_T$ is measured at the system-reservoir boundary over a period \(T \ll L^2\) to minimize boundary effects. The lattice is initially populated following a Bernoulli distribution with uniform average density \(\rho_b\).

The simulation results for the scgf of $Q_T$ shown in Fig. \ref{fig:semi_inf_slow_cgf} are for $T=500$ on a system of length $L=100$ averaged over a clone population $N_c=10^4$.

\subsection{SSEP on finite lattice coupled with two boundary reservoirs}
The scgf in Fig.~\ref{fig: finite_slow_cgf} is for a SSEP on a finite lattice of size \(L\) with a reservoir of density \(\rho_b\) connected at the right end and another reservoir of density \(\rho_a\) at the left end. Schematics is in Fig.~\ref{fig:fin_with_slow_boundary_coupling}. Result in Fig.~\ref{fig: finite_slow_cgf} corresponds to the choice of rates \(\gamma_{a(b)}=\frac{\Gamma}{L}\). The net current is measured at the left boundary over a period \(T \gg L^2\), allowing the system to reach its stationary state.

Implementation of the cloning algorithm is straightforward in this case. The scgf in Fig.~\ref{fig: finite_slow_cgf} corresponds to \((\rho_a, \rho_b) = \left(\frac{1}{2}, \frac{1}{2}\right)\) and a lattice of length \(L = 32\) with a measurement time of \(T = 10^4\) and a clone population of \(N_c = 10^4\).

\section{Conclusion \label{sec:conclusion}}

In this article, we discussed the effect of defect bonds on the large deviation statistics of particle flux for a SSEP on three conventional geometries: finite, semi-infinite, and infinite lattice. These results are based on a fluctuating hydrodynamics framework suitably extended to capture the large scale effects of microscopic defects. We have shown that unless the slow hopping rates across the defect bonds fall below a critical value, both the fluctuating hydrodynamic and the large deviations asymptotics remain unaffected. Specifically, we have derived the scgf of the time-integrated current in each geometry at the critical slow hopping rate. We have further complemented our analytical findings with importance sampling results obtained via a cloning algorithm. 

We showed that the expressions for the scgfs have insightful variational representations (\ref{eq:variational infinite ssep},\ref{eq:variational infinite ssep 2},\ref{eq:var sol fin}) that relate to a prior additivity construction \cite{2021_Derrida_Large} for incorporating effects of localized defect bonds.
This construction, offers an intuitive reasoning for directly obtaining the scgf in presence of defect bonds from the known results in their absence. Notably, the variational construction \eqref{eq:si inf relation} offers a simple, independent derivation of the scgf \eqref{PW_Rsi} on semi-infinite SSEP that earlier required a more elaborate derivation~\cite{2026_Sharma_Large}.

For the SSEP, the scgf of the time-integrated current, regardless of the underlying geometry of the lattice, is known \cite{2009_Derrida_Current} to depend on the fugacity parameter $\lambda$ and the density ($\rho_a$ and $\rho_b$) solely through a single function $\omega(\lambda,\rho_a,\rho_b)$. Within hydrodynamics this is a direct consequence of a rotational symmetry in the MFT action \cite{2009_Derrida_Current2}. Our exact results for different geometries re-affirms that this parametric dependence is robust even in presence of defect bonds.

Although the results presented are for SSEP, they can be extended to systems where the mobility is quadratic in density, such as the symmetric simple inclusion process \cite{2014_Vafayi_Weakly,2022_Franceschini_Symmetric} and the Kipnis-Marchioro-Presutti model \cite{1982_Kipnis_Heat,2007_Giardinà_Duality} of heat transport. For future, it would be interesting to extend the analysis for scenarios where defects are spatially arranged over extended region or moving with time such as in presence of second class of particles. Effects of defect for biased transport models, such as the Asymmetric simple exclusion process (ASEP) will be particularly an interesting problem to study.

The large deviation results reported in this article has interest beyond their appeal from integrability consideration. For example, the scgf of current for SSEP on infnite line gives \cite{2026_Sharma_Large} the time scale for the unusual slow relaxation of spin-auto correlation in a kinetically constrained model \cite{1989_Spohn_Stretched,2024_Mukherjee_Stretched}. Transport in the semi-classical limit of chaotic quantum many-body dynamics is surprisingly captured \cite{2023_McCulloch_Full,turkeshi2024,2026_mcculloch} by the SSEP where these rare event statistics are becoming relevant due to advances in quantum microscopy \cite{2024_Wienand_Emergence}. Understanding the effects of spatial inhomogeneity is particularly relevant for these practical scenarios where our hydrodynamics construction incorporating defects is expected to be useful.

\begin{acknowledgments}
Our work on slow bonds has been deeply influenced by our early collaborations with Bernard Derrida on the finite SSEP coupled to reservoirs through a slow bond. The formulation of the additivity ansatz for the large deviations of the current across a slow bond in the finite SSEP emerged directly from this collaboration. TS gratefully acknowledges many discussions with Bernard Derrida, especially on extending the additivity ansatz to other geometries and dynamical settings. We also acknowledge our collaboration with Thibaut Arnoulx de Pirey, through which the coarse‑graining approach for general stochastic dynamics was developed. We acknowledge financial support from the Department of Atomic Energy, Government of India, under Project Identification Number RTI-4012. The computations were carried out on the computing clusters at the Department of Theoretical Physics, TIFR, Mumbai. We also thank Ajay Salve and Kapil Ghadiali for their computational support. TS further thanks the International Research Project (IRP) “Classical and Quantum Dynamics in Out-of-Equilibrium Systems” funded by CNRS, France, for its support.
\end{acknowledgments}

\bibliography{references}

\end{document}